\documentclass[
  preprint,
  superscriptaddress,
  longbibliography,
  amsmath,
  amssymb,
  aps,
  prx,
  floatfix
]{revtex4-2}

\usepackage[utf8]{inputenc}
\usepackage{graphicx}
\usepackage{bm}
\usepackage{placeins}
\usepackage{xcolor}

\usepackage{hyperref}
\begin{document}

\title{Emergence of Anisotropic Defect Networks in Shocked Diamond}

\author{Zipeng Xu}
\email[Contact author: ]{zipengx@stanford.edu}
\affiliation{Department of Materials Science and Engineering, Stanford University, Stanford, California 94305, USA}
\affiliation{SLAC National Accelerator Laboratory, Menlo Park, California 94025, USA}
\affiliation{PULSE Institute, Stanford University, Stanford, California 94305, USA}

\author{Kento Katagiri}
\affiliation{Department of Materials Science and Engineering, Stanford University, Stanford, California 94305, USA}
\affiliation{SLAC National Accelerator Laboratory, Menlo Park, California 94025, USA}
\affiliation{PULSE Institute, Stanford University, Stanford, California 94305, USA}
\affiliation{Graduate School of Engineering, The University of Osaka, Osaka 565-0871, Japan}
\affiliation{Institute of Laser Engineering, The University of Osaka, Osaka 565-0871, Japan}

\author{Tatiana Pikuz}
\affiliation{Institute for Open and Transdisciplinary Research Initiatives, The University of Osaka, Suita, Osaka 565-0871, Japan}
\affiliation{Institute for Laser Technology, Suita, Osaka 565-0871, Japan}

\author{Wanghui Li}
\affiliation{School of Civil Engineering and Transportation, South China University of Technology, Guangzhou 510640, China}
\affiliation{Institute of High Performance Computing (IHPC), Agency for Science, Technology and Research (A*STAR), Singapore 138632, Republic of Singapore}

\author{Anirudh Hari}
\affiliation{Department of Materials Science and Engineering, Stanford University, Stanford, California 94305, USA}
\affiliation{SLAC National Accelerator Laboratory, Menlo Park, California 94025, USA}
\affiliation{PULSE Institute, Stanford University, Stanford, California 94305, USA}

\author{Dorian P. Luccioni}
\affiliation{Department of Materials Science and Engineering, Stanford University, Stanford, California 94305, USA}
\affiliation{SLAC National Accelerator Laboratory, Menlo Park, California 94025, USA}
\affiliation{PULSE Institute, Stanford University, Stanford, California 94305, USA}

\author{Sophie E. Parsons}
\affiliation{Department of Materials Science and Engineering, Stanford University, Stanford, California 94305, USA}
\affiliation{SLAC National Accelerator Laboratory, Menlo Park, California 94025, USA}
\affiliation{PULSE Institute, Stanford University, Stanford, California 94305, USA}

\author{Yuichi Inubushi}
\affiliation{Japan Synchrotron Radiation Research Institute, Sayo 679-5198, Japan}
\affiliation{RIKEN SPring-8 Center, Sayo 679-5148, Japan}

\author{Kohei Miyanishi}
\affiliation{RIKEN SPring-8 Center, Sayo 679-5148, Japan}

\author{Hirotaka Nakamura}
\affiliation{Graduate School of Engineering, The University of Osaka, Osaka 565-0871, Japan}

\author{Tadashi Togashi}
\affiliation{Japan Synchrotron Radiation Research Institute, Sayo 679-5198, Japan}
\affiliation{RIKEN SPring-8 Center, Sayo 679-5148, Japan}

\author{Makina Yabashi}
\affiliation{Japan Synchrotron Radiation Research Institute, Sayo 679-5198, Japan}
\affiliation{RIKEN SPring-8 Center, Sayo 679-5148, Japan}

\author{Toshinori Yabuuchi}
\affiliation{Japan Synchrotron Radiation Research Institute, Sayo 679-5198, Japan}
\affiliation{RIKEN SPring-8 Center, Sayo 679-5148, Japan}

\author{Norimasa Ozaki}
\affiliation{Graduate School of Engineering, The University of Osaka, Osaka 565-0871, Japan}
\affiliation{Institute of Laser Engineering, The University of Osaka, Osaka 565-0871, Japan}

\author{Leora E. Dresselhaus-Marais}
\email[Contact author: ]{leoradm@stanford.edu}
\affiliation{Department of Materials Science and Engineering, Stanford University, Stanford, California 94305, USA}
\affiliation{SLAC National Accelerator Laboratory, Menlo Park, California 94025, USA}
\affiliation{PULSE Institute, Stanford University, Stanford, California 94305, USA}
\date{November 13, 2026}

\begin{abstract}
Predicting failure after shock loading requires understanding how the structures created during compression evolve as the material unloads. In diamond, both crystal plasticity and localized amorphization have been reported to accommodate shock-induced shear. However, it remains unresolved how the resulting defect structures organize in three dimensions during compression and how this organization influences where tensile damage localizes during release. Here we couple phase-contrast-enhanced radiography at an X-ray free-electron laser with simulations to address this question in single-crystal diamond. By combining radiographs at multiple pump-probe delays with hydrodynamic calculations, we show that linear features first become visible behind the plastic wavefront while material near the front remains highly compressed. By rotating the viewing direction about the shock axis, we find that the projected feature angles favor finite $\{111\}$ sheets elongated along in-plane directions close to $\langle112\rangle$. Molecular dynamics simulations show that an interconnected network of amorphous bands and stacking faults forms during compression, preferentially on $\{111\}$ variants with larger in-plane shear traction. Projections of this network qualitatively reproduce the observed loss of feature visibility with rotation. During simulated release, tension opens local cracks within these pre-existing amorphous bands. These results support a pathway in which the crystallographically organized defect network formed during compression provides preferred sites for tensile damage during release. This pathway may also operate in other strongly covalent solids.
\end{abstract}

\maketitle

\section{Introduction}
Predicting failure after shock loading requires an understanding of how the structures created during compression evolve as the material unloads. In strongly covalent solids, where high lattice resistance limits dislocation motion, crystal plasticity and localized amorphization can both accommodate deformation~\cite{Zhao2021Directional}. For example, in boron carbide, shock-induced amorphous bands aligned with particular crystallographic planes border fracture surfaces and have been linked to its loss of ballistic performance at high impact pressures~\cite{Chen2003Shock}. A central question is how crystalline defects and locally disordered regions develop into extended structures during shock compression, and how the orientation and morphology of those structures influence where tensile damage localizes during release. Diamond provides a model system for addressing this question. It shares tetrahedral bonding with silicon and silicon carbide and sustains exceptionally large, orientation-dependent stresses~\cite{McWilliams2010Strength,Lang2018Strength,Winey2020Shock}. Establishing the connection between these deformation structures and subsequent damage would provide microstructural constraints for models of dynamic strength and failure.

Wave-profile measurements and time-resolved X-ray diffraction have quantified diamond's elastic--plastic response and strength across different crystal orientations and microstructures~\cite{McWilliams2010Strength,Lang2018Strength,Winey2020Shock,MacDonald2020Using,Hari2026Ultrahigh}. Connecting this mechanical response to its microscopic origins requires characterizing the three-dimensional morphology and evolution of the associated defect structures. Microscopy and atomistic simulations have found both crystal plasticity and localized amorphization in diamond under different loading conditions and at different length scales~\cite{Li2025Direct,Nie2020Dislocation,Zhang2026NanodiamondPlasticity,Li2023Dislocation}. In recovered laser-shocked diamond, narrow amorphous bands coexist with dislocations and stacking faults, and the onset of amorphization has been linked to shear-induced lattice defects~\cite{Li2025Direct}. However, recovered microstructures record the outcome of both compression and release; they do not distinguish when these structures formed or how they changed during unloading. Establishing the deformation pathway is necessary to connect the observed defects with the tensile fracture known to occur during release in diamond~\cite{Abrosimov2014Spallation}.

Imaging at X-ray free-electron lasers (XFELs) can constrain this pathway through spatially resolved observations at multiple pump-probe delays. Vennari et~al.~\cite{Vennari2026Topography} developed single-pulse X-ray topography to probe lattice distortions in ambient and shock-compressed diamond through diffraction contrast imaging. Phase-contrast-enhanced radiography has resolved elastic-wave propagation and elastic--plastic wave splitting in diamond~\cite{Schropp2015Imaging,Makarov2023Direct}. Katagiri et~al.~\cite{Katagiri2023Transonic} observed inclined linear features behind the plastic wavefront, associated their orientations with $\{111\}$ planes, and interpreted their extension in terms of stacking-fault growth driven by transonic partial-dislocation motion. A comparison of two probing directions in the same study further supported the planar character of these features. However, their crystallographic alignment alone does not establish the three-dimensional morphology of the underlying structures or uniquely distinguish stacking faults from other planar structures, such as amorphous bands or cracks~\cite{Li2025Direct,Hawreliak2024Comments,Katagiri2024Comment}. It is also not clear how those linear features evolve during release and relate to subsequent damage and fragmentation.

Here we analyze phase-contrast-enhanced radiographs spanning multiple pump-probe delays and viewing directions. The delay series and hydrodynamic calculations show that the linear features emerge in highly compressed material behind the plastic wavefront. As the viewing direction is rotated, the projected angles of the features shift and their visibility decreases. These angle shifts favor finite $\{111\}$ sheets elongated along in-plane directions close to $\langle112\rangle$. A large-scale shock--release molecular dynamics (MD) simulation produces an interconnected network of amorphous bands and stacking faults on the $\{111\}$ variants with larger in-plane shear traction; projections of this network qualitatively reproduce the loss of visibility with rotation. During simulated release, local cracks open within the pre-existing amorphous bands under tension. Together, these results support a mechanism in which the crystallographic organization of the defect network formed during compression biases the localization of tensile damage during release.

\section{Methods}
\subsection{Laser-shock compression and XFEL radiography}

The laser-shock/XFEL experiments were performed at the SPring-8 Angstrom Compact Free Electron Laser (SACLA) facility at BL3-EH5 using the pump-probe configuration described in Refs.~\cite{Makarov2023Direct,Katagiri2023Transonic}. Each target consisted of a single-crystal diamond sample bonded to a 50~$\mu$m-thick polypropylene (PP) ablator. The diamond samples measured $2.0\times0.5~\mathrm{mm}^2$, with a thickness of 200--300~$\mu$m along the shock direction depending on the crystal orientation. We used type IIa single-crystal diamonds grown by chemical vapor deposition (CVD) and purchased from EDP Corporation (Japan), with shock propagation along either the $[100]$ or $[110]$ direction. The shock was driven by a 532-nm laser pulse with a 5-ns flat-top temporal profile. A diffractive optical element (DOE) shaped the transverse intensity distribution to produce an approximately flat-top focal spot with a diameter of 260~$\mu$m on the target.

The diamond was probed perpendicular to the shock direction using unfocused, 7.0~keV XFEL pulses. The XFEL pulse energy was approximately 0.34~mJ, the pulse duration was approximately 8~fs, and the beam diameter at the sample was approximately 800~$\mu$m~\cite{Katagiri2023Transonic}. Phase-contrast-enhanced radiographs were recorded using 0.9~mm-thick LiF crystal detectors with their front surfaces positioned 112~mm downstream of the sample. The LiF images were read out by confocal fluorescence microscopy~\cite{Pikuz2018LiF} at $4\times$, $10\times$ and $20\times$ magnification, with spatial sampling of 1.55, 0.62 and 0.31~$\mu$m per pixel, respectively.

The recorded intensity contains both attenuation and phase-contrast contributions associated with the material distribution along the X-ray path. With sufficient spatial coherence, Fresnel diffraction over the finite sample--detector distance converts transverse variations in the transmitted phase into intensity modulations~\cite{Snigirev1995Phase,Wilkins1996Phase,Pogany1997Contrast}. At 7.0~keV ($\lambda\approx0.177$~nm), the sample--detector distance of $z=112$~mm corresponds to a characteristic Fresnel scale $\ell_{\mathrm F}=\sqrt{\lambda z}\approx4.5~\mu$m. For hard X-rays in diamond, the refractive-index decrement is much larger than the absorptive component of the refractive index, making propagation-based phase contrast well suited to imaging structures with weak attenuation contrast~\cite{Schropp2015Imaging}.

Two types of radiographic datasets were analyzed: pump-probe delay series and rotation series at a fixed delay. Unless otherwise noted, pump-probe delays are referenced to the arrival of the drive laser at the front surface of the PP ablator. Each radiograph corresponds to an independent shot.

The $[100]$ delay series comprised 16 independent laser shots at pump-probe delays of 4--1000~ns, with an average on-target drive-pulse energy of $15.79\pm1.34$~J. Selected radiographs in the $[100]$ delay series are based on original LiF records from shots reported in Fig.~S12 of Ref.~\cite{Katagiri2023Transonic}. These records are analyzed here together with additional measurements.

The $[110]$ rotation series comprised 19 shots at a fixed pump-probe delay of 12~ns, covering $\phi=0^\circ$--$90^\circ$ in $5^\circ$ increments, with an average on-target drive-pulse energy of $16.91 \pm 1.04$~J. The crystal was rotated about the fixed $[110]$ shock axis, with $\phi=0^\circ$ corresponding to an XFEL line of sight parallel to $[\bar{1}10]$. Quantitative rotation-dependent analyses were restricted to $\phi\leq40^\circ$. At $\phi\geq45^\circ$, the increased X-ray path length through the sample reduced the transmitted intensity, and the two families of linear features could not be resolved reliably.

An additional $[100]$ rotation series was acquired at a fixed pump-probe delay of 12~ns, covering rotations from $0^\circ$ to $45^\circ$ about the fixed $[100]$ shock axis. The average on-target drive-pulse energy was $16.55\pm1.55$~J. This series is presented in Supplemental Fig.~S3.

\subsection{Shock--release molecular dynamics simulation}
\label{sec:md-methods}

MD simulations have been used to resolve shock-induced deformation and spallation mechanisms in covalent ceramics and high-entropy alloys~\cite{Li2017SiCSpall,Li2019SiCDamage,Li2023HEASpall}. Here, a shock--release MD simulation of an initially defect-free diamond crystal was performed using LAMMPS~\cite{thompson2022lammps,plimpton1995fast}. The diamond crystal was generated with Atomsk~\cite{hirel2015atomsk}, with the $z$ axis parallel to the $[110]$ shock direction. The simulation cell contained approximately $1.7\times10^{8}$ carbon atoms with dimensions of about $820\times820\times1518~\text{\AA}^{3}$. 

Interatomic interactions were described using the quadratic spectral neighbor analysis potential (SNAP) for carbon \cite{thompson2015snap,willman2022carbon}. This potential was developed and validated for carbon over a broad range of pressures and temperatures relevant to extreme-condition applications, and its performance has subsequently been benchmarked against alternative carbon potentials under similar conditions \cite{willman2024benchmark}. The same potential framework has also been employed in recent large-scale simulations of compressed diamond and post-diamond transformation pathways \cite{nguyencong2024bc8}.

The crystal was equilibrated at 300~K and zero pressure before loading. During shock propagation, periodic boundary conditions were retained in the transverse directions, while nonperiodic boundaries were used along the $[110]$ loading direction. A planar piston wall at the lower $z$ boundary was driven at a constant speed of $u_p=4.5~\mathrm{km/s}$ for 6~ps. This speed was selected to generate an elastic--plastic two-wave response qualitatively similar to that observed experimentally. Simulation time $t=0$ corresponds to the onset of piston motion. At $t=6$~ps, the piston wall was removed, and the subsequent release evolution was analyzed up to $t=15.5$~ps.

One-dimensional profiles of virial stress~\cite{thompson2009stress}, particle velocity, and transverse temperature were obtained by averaging over slabs normal to the loading direction. Transverse temperature was evaluated from the transverse velocity components to exclude streaming motion along the loading direction. Atomic configurations were visualized and structurally analyzed using OVITO~\cite{stukowski2010ovito}. Local atomic environments were classified using polyhedral template matching (PTM), which provides robust structural identification in the presence of thermal fluctuations and lattice distortions~\cite{Larsen2016PTM}.
Additional details of simulation initialization and integration procedures are provided in Supplemental Material, Sec.~II.

\section{Results}

The results begin with the $[100]$ delay series, which establishes the temporal ordering of elastic--plastic wave splitting, linear-feature appearance, and changes in sample morphology after shock breakout. Here, linear features denote narrow, elongated contrast features behind the plastic wavefront in experimental radiographs and MD defect projections; in the reference views, they form two oppositely inclined families. The $[110]$ rotation series then characterizes their viewing-direction dependence. The MD results describe the wave structure and local atomic environments during compression and release. Finally, selected experimental radiographs and MD defect projections are compared in real and Fourier space.

\subsection{Elastic--plastic wave splitting and emergence of linear features}
\label{sec:wave-splitting-linear-features}

\begin{figure}
    \centering
    \includegraphics[width=1.0\textwidth]{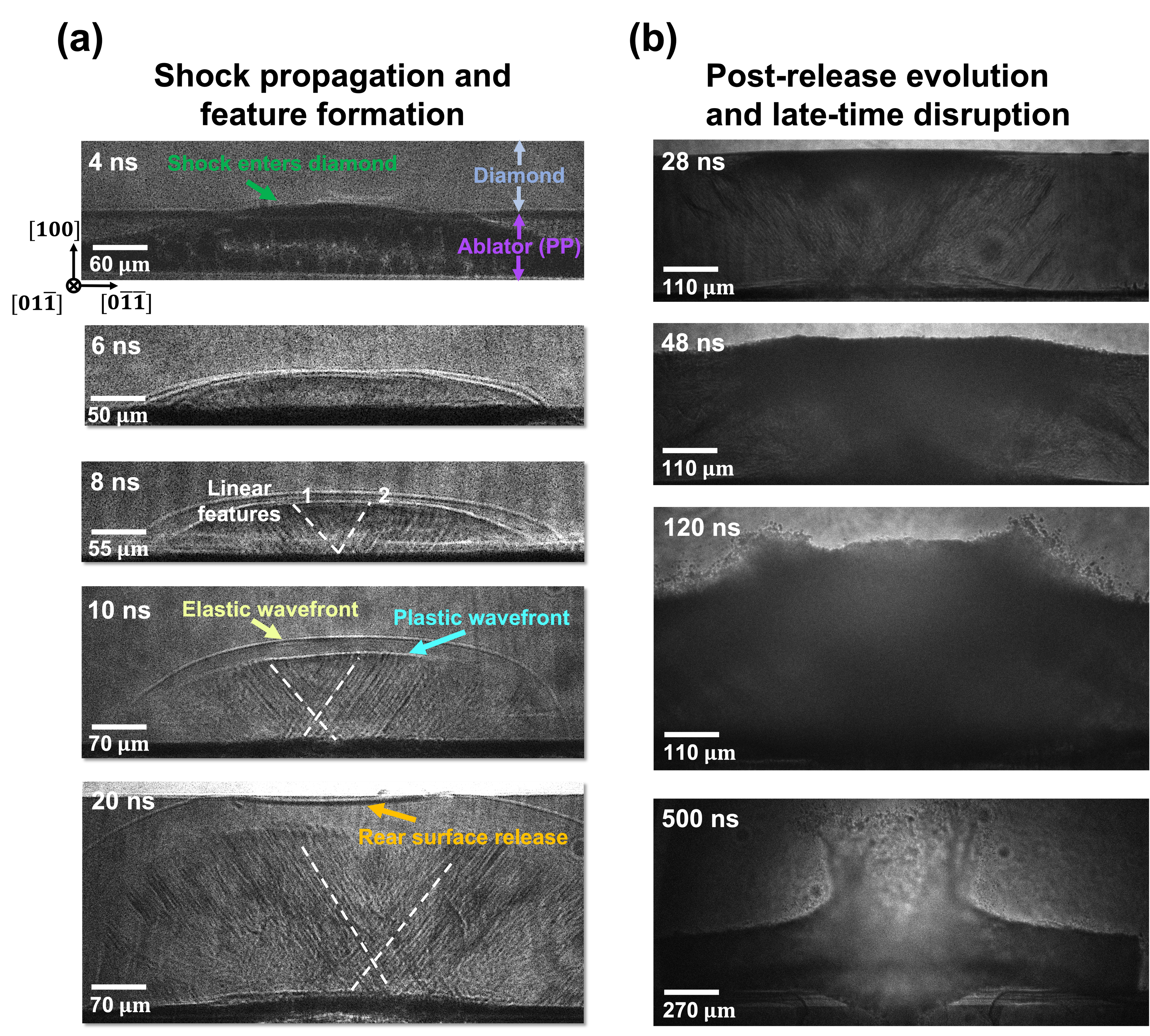}
    \caption{\textbf{Elastic--plastic wave splitting and linear-feature evolution in laser-shocked $[100]$ single-crystal diamond.} (\textbf{a}) Radiographs at pump-probe delays of 4, 6, 8, 10, and 20~ns. Arrows identify the elastic and plastic wavefronts and the rear-surface release front; dashed guides mark the two dominant linear-feature families, Families~1 and~2.
    (\textbf{b}) Radiographs at 28, 48, 120, and 500~ns. LiF readout magnifications were $20\times$ for (a) and $10\times$ for (b), except at 500~ns ($4\times$). 
    Each image represents an independent laser shot and was cropped, annotated, and rescaled with its own scale bar. 
    The radiographs shown here were prepared from the original LiF records of selected shots reported in Fig.~S12 of Ref.~\cite{Katagiri2023Transonic}. Previously published material modified from K.~Katagiri et al., \textit{Science} \textbf{382}, 69--72, doi:10.1126/science.adh5563 (2023), AAAS.}
    \label{fig:fig1}
\end{figure}

Figure~\ref{fig:fig1} presents selected radiographs from the $[100]$ delay series. At 4~ns (Fig.~\ref{fig:fig1}(a)), the dark boundary separating the two rectangular regions of different intensity marks the PP--diamond interface. A curved dark feature extends through the PP ablator and slightly into the diamond, consistent with shock entry at this delay.

At 6~ns, two closely spaced curved fronts are resolved in the diamond and identified as the elastic precursor and trailing plastic wavefront (light yellow and cyan arrows, respectively, in the 10-ns panel of Fig.~\ref{fig:fig1}(a)). The separation between the two wavefronts is larger at 10~ns than at 6~ns, consistent with elastic--plastic wave splitting.
At 20~ns, an additional curved feature (denoted by the orange arrow) appears near the rear free surface and extends into the sample interior. Its position and appearance after shock breakout are consistent with a release front originating at the rear free surface.

Axial wavefront positions were measured from the initial PP--diamond interface in each shot's pre-shot radiograph. Their variation with pump-probe delay gives propagation speeds of $17.3\pm1.1~\mathrm{km/s}$ for the elastic precursor and $14.0\pm1.2~\mathrm{km/s}$ for the plastic wavefront, and neither front shows appreciable slowing over the measured propagation interval.

Linear features are first clearly resolved at 8~ns and extend throughout the imaged region behind the plastic wavefront, forming two oppositely inclined families (Families~1 and~2, labeled in the 8-ns panel of Fig.~\ref{fig:fig1}(a)). At 10~ns, the plastic wavefront has advanced farther into the sample, and both families extend across the enlarged region behind it. Their signed projected angles $\beta$, measured from the detector horizontal (parallel to $[0\bar{1}\bar{1}]$), are $(-51.6\pm1.3)^\circ$ for Family~1 and $(+52.4\pm2.4)^\circ$ for Family~2. Both families remain resolvable at 20~ns, and their projected orientations are approximately unchanged from 8 to 20~ns. Unless otherwise noted, projected orientations are reported as the absolute angle $|\beta|$. Experimental feature-family angles were obtained by line-segment analysis of Fourier-filtered radiographs, and the reported uncertainties represent standard deviations. The extraction procedure is described in Supplemental Material, Sec.~I, with a representative analysis in Supplemental Fig.~S1.

Figure~\ref{fig:fig1}(b) shows later radiographs following rear-surface release. At 28~ns, both feature families remain resolvable across much of the sample, while spatially diffuse contrast appears near the rear free surface. By 48~ns, diffuse contrast dominates the image, and individual linear features are no longer clearly resolved, although a weak remnant of the two families remains visible. At 120~ns, the contrast remains diffuse and sample swelling is apparent. By 500~ns, the sample has fragmented and its original geometry is no longer preserved.

\subsection{Viewing-direction dependence of linear contrast in diamond shocked along [110]}
\label{sec:viewing-direction-dependence}

The $[100]$ delay series establishes when the linear features appear relative to wave splitting (Sec.~\ref{sec:wave-splitting-linear-features}). The $[110]$ rotation series examines how their projected angles and visibility change as the crystal is rotated about the shock axis at a fixed pump-probe delay of 12~ns. The $[110]$ orientation was selected because the linear features showed clearer angular signatures than under $[100]$ loading (Supplemental Fig.~S3).

\begin{figure}
	\centering
	\includegraphics[width=1.0\textwidth]{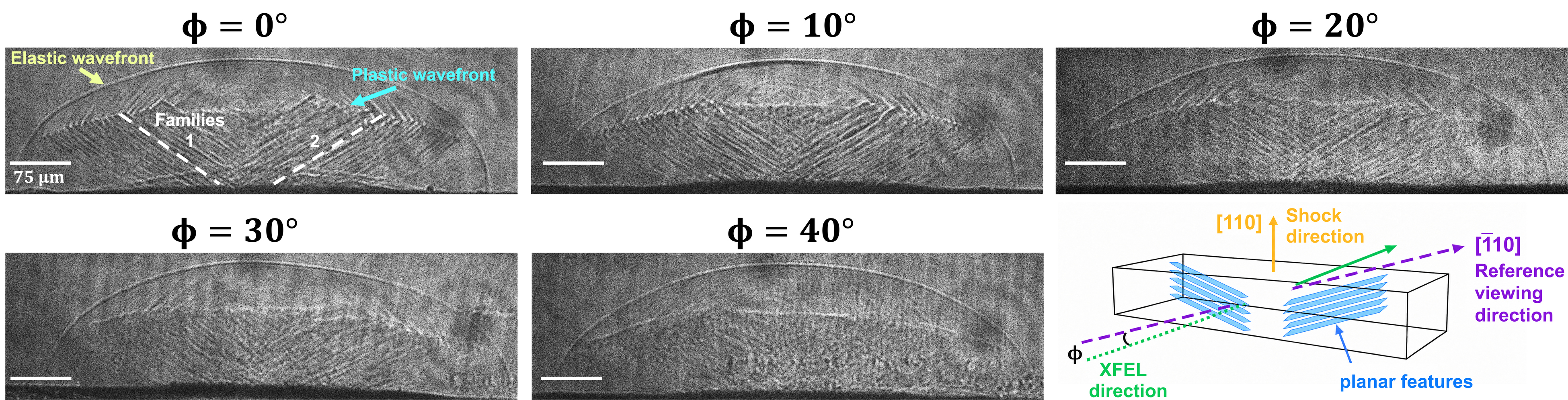}

	\caption{\textbf{Viewing-direction dependence of linear contrast in single-crystal diamond shocked along $[110]$.}
	Selected XFEL radiographs acquired at a pump-probe delay of 12~ns are shown for crystal rotations about the fixed $[110]$ shock axis. 
    In the $\phi=0^\circ$ radiograph, light-yellow and cyan arrows mark the elastic and plastic wavefronts, respectively; white dashed guides mark Families~1 and~2. 
    The schematic shows the viewing geometry, with internal features represented as planes. The rotation angle $\phi$ is measured from the reference XFEL viewing direction $[\bar{1}10]$. 
    Scale bars, 75~$\mu$m.}
	\label{fig:fig2}
\end{figure}

Figure~\ref{fig:fig2} presents five representative radiographs from this series. Here, $\phi$ denotes the crystal rotation about the $[110]$ shock axis, with $\phi=0^\circ$ corresponding to an XFEL line of sight parallel to $[\bar{1}10]$. At $\phi=0^\circ$, two oppositely inclined feature families form a well-resolved intersecting pattern behind the plastic wavefront. Families~1 and~2 have mean absolute projected angles of $(33.2\pm1.9)^\circ$ and $(32.0\pm1.4)^\circ$, respectively. Individual linear features remain resolved at $\phi=10^\circ$ and $20^\circ$. At $\phi=30^\circ$, the directional contrast is weaker and more diffuse than at $\phi=0^\circ$. At $\phi=40^\circ$, individual linear features are no longer resolved.

We next use a shock--release MD simulation to examine the structures generated during $[110]$ compression and their evolution during release.

\subsection{Wave structure and evolution of local atomic environments during shock compression and release}
\label{sec:md-wave-structure}

The MD simulation follows single-crystal diamond shocked along $[110]$ and viewed along $[\bar{1}10]$, corresponding to the reference geometry of the rotation experiment at $\phi=0^\circ$. The piston moves at $u_p=4.5~\mathrm{km/s}$ until 6~ps and is then removed. The simulation continues to 15.5~ps to follow the subsequent release.

\begin{figure}
    \centering
    \includegraphics[width=0.9\textwidth]{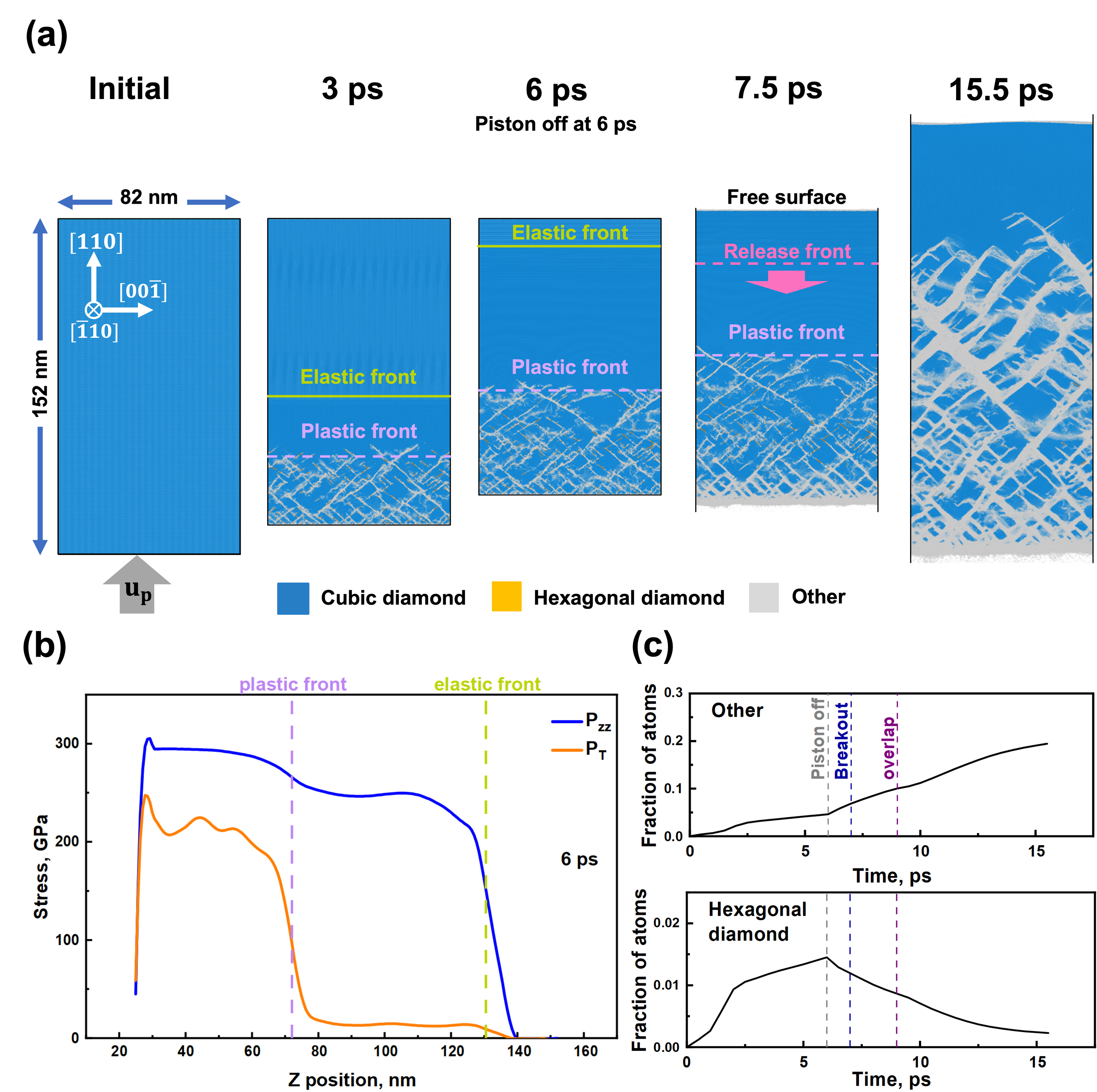}
    \caption{\textbf{Wave structure and local atomic environments during shock compression and release in diamond shocked along $[110]$.}
    (\textbf{a}) Initial configuration and local structure maps at 3, 6, 7.5, and 15.5~ps. The initial dimensions shown are 82~nm along $[00\bar{1}]$ and 152~nm along $[110]$; the viewing direction is $[\bar{1}10]$. The piston moves at $u_p=4.5~\mathrm{km/s}$ until 6~ps and is then removed. Blue, yellow, and gray denote Cubic diamond, Hexagonal diamond, and Other, respectively. Lines mark the wavefronts; the arrow at 7.5~ps indicates the release-front propagation direction.
    (\textbf{b}) Longitudinal and mean transverse compressive stresses at 6~ps, with $P_{zz}=-\sigma_{zz}$ and $P_T=-(\sigma_{xx}+\sigma_{yy})/2$.
    (\textbf{c}) Fractions of atoms classified as Other and Hexagonal diamond in the entire simulated crystal versus time. Vertical dashed lines mark piston shutoff, elastic-wave breakout at the upper free surface, and overlap of the rear-surface release front with the plastic wavefront.}
    \label{fig:fig3}
\end{figure}

Each atom is classified as Cubic diamond, Hexagonal diamond, or Other according to its local atomic environment. Before loading, the bulk crystal is classified entirely as Cubic diamond (Fig.~\ref{fig:fig3}(a)). By 3~ps, the elastic and plastic fronts have separated, and their separation increases to approximately 60~nm at 6~ps (see Supplemental Material, Sec.~VII, for wavefront identification). At 6~ps, the unshocked, elastically compressed, and plastically compressed regions are clearly resolved. The 6~ps configuration, at the end of piston driving, serves as the compression-stage reference for the analyses below.

The stress profiles at 6~ps quantify the elastic and plastic two-wave state (Fig.~\ref{fig:fig3}(b)). Moving from the unshocked material toward the piston, $P_{zz}$ rises across the elastic front to approximately $240$--$250~\mathrm{GPa}$, while $P_T$ remains small relative to $P_{zz}$ in the region between the two fronts. Across the plastic front, $P_T$ rises toward approximately $210~\mathrm{GPa}$, while $P_{zz}$ increases further to approximately $290~\mathrm{GPa}$. 

Changes in local atomic environments occur primarily behind the plastic front. At 3 and 6~ps, atoms classified as Other form two oppositely inclined families of narrow features connected to segments classified as Hexagonal diamond (Fig.~\ref{fig:fig3}(a)). Together, these features form an intersecting noncubic network, while the region between the fronts remains predominantly Cubic diamond. The noncubic network occupies a larger region as the plastic front advances from 3 to 6~ps.

At 7.5~ps, a release front propagates from the upper free surface back into the compressed crystal, while the noncubic network remains clearly resolved behind the plastic front (Fig.~\ref{fig:fig3}(a)). By 15.5~ps, the individual features have broadened, and the network is more interconnected than at 6~ps. Atoms classified as Other occupy most of the broadened regions, while segments classified as Hexagonal diamond remain visible in parts of the crystal.

The fractions of atoms classified as Other and Hexagonal diamond, calculated over the entire simulated crystal, follow different trends (Fig.~\ref{fig:fig3}(c)). The Other fraction increases during the driven stage and continues to rise after piston shutoff, reaching approximately 0.19 at 15.5~ps. The Hexagonal diamond fraction peaks at approximately 0.014 at 6~ps and then decreases to approximately 0.002 by 15.5~ps.

\subsection{Comparison of experimental radiographs and MD defect projections}

\begin{figure}
    \centering
    \includegraphics[width=1.0\textwidth]{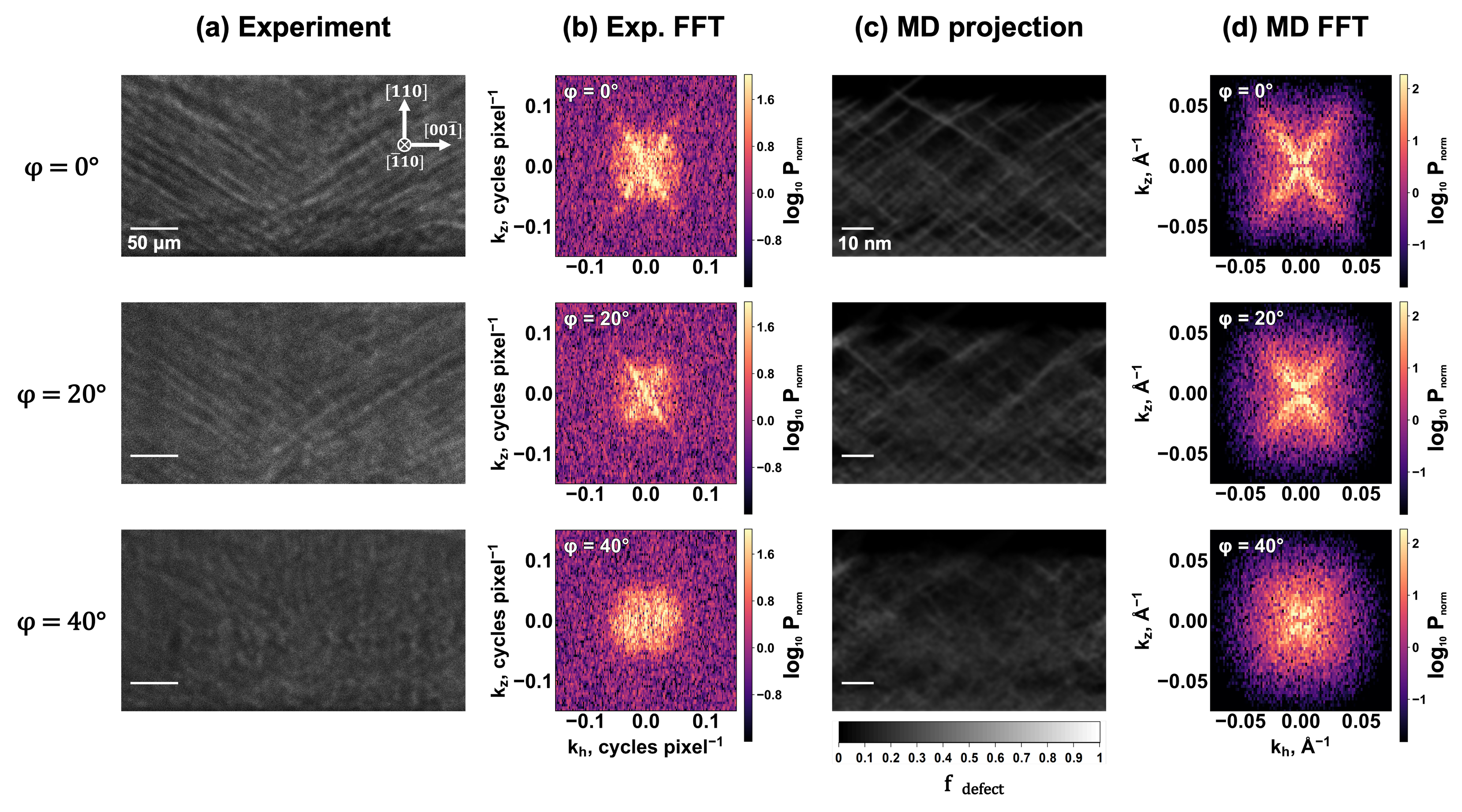}
    \caption{\textbf{Experimental radiographs, MD defect projections, and corresponding Fourier power spectra for diamond shocked along $[110]$.}
    Rows correspond to $\phi=0^\circ$, $20^\circ$, and $40^\circ$ from top to bottom, where $\phi$ denotes rotation about the $[110]$ shock axis. At $\phi=0^\circ$, the viewing direction is parallel to $[\bar{1}10]$ and the image horizontal direction is parallel to $[00\bar{1}]$.
    (\textbf{a}) Selected regions of phase-contrast-enhanced radiographs acquired at a pump-probe delay of 12~ns.
    (\textbf{b}) Corresponding two-dimensional Fourier power spectra of the experimental radiographs.
    (\textbf{c}) Projected defect-fraction maps from the 6~ps MD configuration in Fig.~\ref{fig:fig3}(a). The common grayscale denotes the fraction of atoms classified as Other and Hexagonal diamond along each projected line of sight, $f_{\mathrm{defect}}$.
    (\textbf{d}) Corresponding Fourier power spectra of the MD projections. Colors in (\textbf{b,d}) indicate the logarithmic normalized Fourier power. The horizontal and vertical spatial-frequency components, $k_{\mathrm{h}}$ and $k_z$, are expressed in $\mathrm{cycles\,pixel^{-1}}$ in (\textbf{b}) and $\text{\AA}^{-1}$ in (\textbf{d}). Scale bars are $50~\mu\mathrm{m}$ in (\textbf{a}) and $10~\mathrm{nm}$ in (\textbf{c}) and apply to all three images in the respective columns.}
    \label{fig:fig4}
\end{figure}

We compare the 12~ns experimental radiographs with defect-fraction projections of the 6~ps MD configuration at matched crystallographic viewing directions. The experiment and simulation differ in spatial scale and loading state, and the radiographs and MD projections represent X-ray intensity and projected defect fraction, respectively. The comparison therefore focuses on the morphology and directional organization of the feature families and their variation with viewing direction, without assuming one-to-one correspondence between individual features or a quantitative mapping between the two contrast measures. The 6~ps configuration corresponds to the end of the piston-driven stage, when the defect network behind the plastic front is well resolved (Fig.~\ref{fig:fig3}(a)). Rotating this fixed configuration about the $[110]$ shock axis through the experimental angles $\phi$ isolates the effect of viewing geometry (see Supplemental Material, Sec.~III, for projection details).

Figure~\ref{fig:fig4}(a,c) compares the experimental radiographs and MD projections at $\phi=0^\circ$, $20^\circ$, and $40^\circ$. At $\phi=0^\circ$, two oppositely inclined families of linear features are resolved in both datasets. At $\phi=20^\circ$, the intersecting pattern remains visible, although individual linear features are less sharply defined than at $\phi=0^\circ$. At $\phi=40^\circ$, individual linear features are no longer clearly resolved in the experimental radiograph. The MD projection is likewise more diffuse, retaining only weak oriented contrast.

The angular distribution of Fourier power provides a common basis for comparing the directional organization of the two datasets. At $\phi=0^\circ$, both datasets exhibit enhanced power along two oblique directions (Fig.~\ref{fig:fig4}(b,d)). At $\phi=20^\circ$, these directional features remain visible but are less distinct against the spectral background. By $\phi=40^\circ$, the two-direction pattern has weakened further in both datasets. The full set of Fourier spectra at $5^\circ$ intervals from $\phi=0^\circ$ to $40^\circ$ shows an overall weakening of the oblique directional features with increasing rotation angle (Supplemental Fig.~S4). This trend accompanies the decreasing visibility of individual linear features in real space.

Taken together, the experimental radiographs and MD projections show similar directional organization and a similar dependence on viewing direction, despite their different spatial scales.

\section{Discussion}

We now examine the loading state in which the linear features emerge, the identity of the underlying structures and their evolution during release, and the three-dimensional morphology implied by their rotation dependence.

\subsection{Loading and release history of the linear features}
\label{sec:linear-feature-timing}

\begin{figure}
    \centering
    \includegraphics[width=1.0\textwidth]{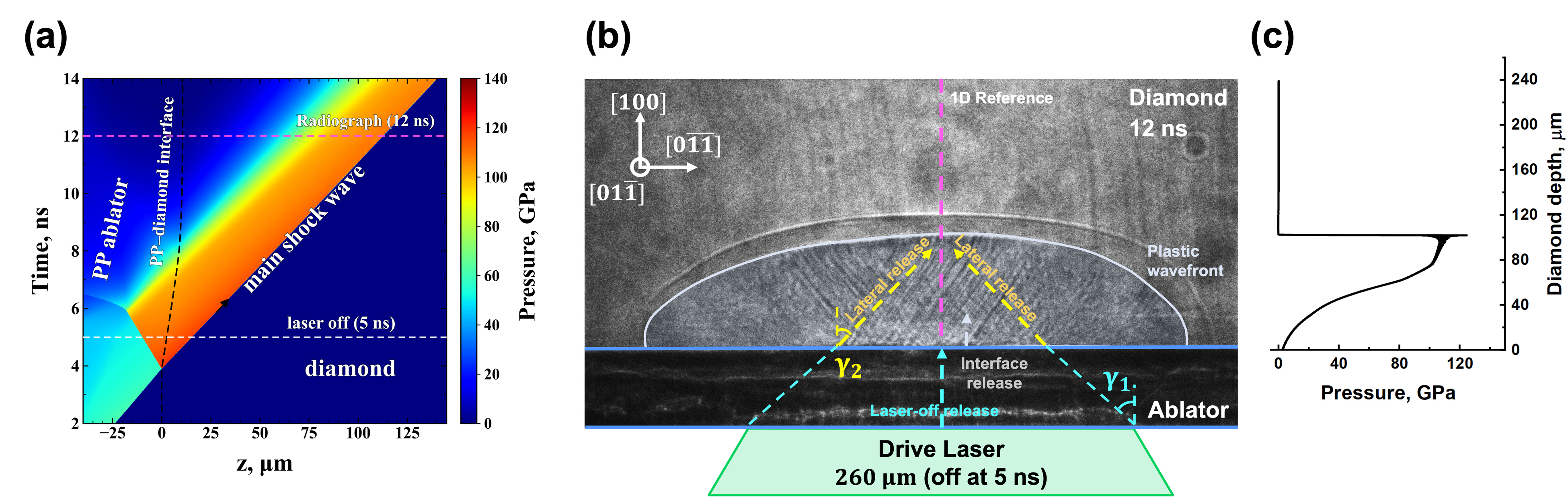}
    \caption{\textbf{Loading and release in the $[100]$ diamond target.}
    (\textbf{a}) Position--time pressure map from the one-dimensional
    HELIOS calculation calibrated to the $[100]$ shock-entry time
    and plastic wavefront propagation speed. The coordinate $z$
    is measured from the initial PP--diamond interface. The black
    dashed curve follows the moving interface. The white and
    magenta horizontal dashed lines mark laser termination at
    5~ns and the radiographic delay of 12~ns, respectively.
    (\textbf{b}) Phase-contrast-enhanced radiograph acquired at
    12~ns, with release directions and the calculated shock-erosion
    construction overlaid. The cyan and gray axial arrows indicate
    the propagation directions of laser-off release and interface
    release, respectively. The cyan and yellow inclined dashed
    lines show the shock-erosion construction in the PP and
    diamond, respectively. In diamond, the construction is applied
    to the plastic wavefront. The angles $\gamma_1$ and $\gamma_2$
    are measured relative to the shock-propagation direction.
    The upper blue horizontal line marks the PP--diamond interface,
    and the white curve follows the plastic wavefront. The magenta
    dashed line marks the shock-propagation axis used as the
    one-dimensional reference.
    (\textbf{c}) Calculated pressure profile in diamond at 12~ns,
    extracted from (\textbf{a}). Depth is measured from the current
    PP--diamond interface.}
    \label{fig:fig5}
\end{figure}

Linear features are first clearly resolved behind the plastic wavefront at 8~ns and extend farther into the diamond by 10~ns (Fig.~\ref{fig:fig1}(a)). Both observations occur after the 5~ns drive pulse has ended, while the elastic precursor and plastic wavefront continue to propagate and release waves travel from the drive side. To relate these observations to the local pressure history, we performed a one-dimensional planar HELIOS calculation without material strength for a 50-$\mu$m PP ablator and a 250-$\mu$m diamond layer~\cite{MacFarlane2006HELIOS}. The effective drive amplitude was selected to match the approximate 4~ns shock-entry time and the measured $[100]$ plastic wavefront speed of $14~\mathrm{km/s}$. The calculation gives shock entry into the diamond at approximately 3.9~ns and a speed of $14.0~\mathrm{km/s}$ for the main hydrodynamic compression front, which serves as the propagation reference for the experimental plastic wavefront. Figure~\ref{fig:fig5}(a) shows the resulting position--time pressure map from 0 to 14~ns.

The calculation shows that pressure remains high immediately behind the advancing compression front through 14~ns. The local peak pressure decreases from approximately $136~\mathrm{GPa}$ shortly after shock entry to approximately $120~\mathrm{GPa}$ at 12~ns, a reduction of about 10\%. The appearance and extension of the linear features at 8--10~ns therefore occur while material near the plastic wavefront remains highly compressed. At the same time, material farther behind the front, near the PP--diamond interface, progressively unloads. This drive-side unloading includes \emph{laser-off release}, generated by laser termination at 5~ns, and \emph{interface release}, associated with shock reflection and subsequent wave interactions in the PP--diamond system~\cite{Swift2008PlasticAblators}. The axial arrows in Fig.~\ref{fig:fig5}(b) indicate the propagation directions of these release waves. At 12~ns, the calculated pressure profile rises from approximately $3~\mathrm{GPa}$ near the PP--diamond interface to approximately $120~\mathrm{GPa}$ immediately behind the compression front (Fig.~\ref{fig:fig5}(c)), showing the coexistence of drive-side unloading and high compression near the front.

The one-dimensional HELIOS calculation describes axial loading and drive-side unloading. To evaluate lateral release from the edges of the laser spot, we estimate the inward erosion of the plastic wavefront using the shock-erosion construction~\cite{Swift2008PlasticAblators} shown in Fig.~\ref{fig:fig5}(b). With the orientation-dependent sound speeds, lateral erosion is estimated to reach the plastic wavefront on the shock axis at approximately 12~ns for $[100]$ and 9~ns for $[110]$. The erosion angles $\gamma_1$ and $\gamma_2$, the associated transit estimates, and the rear-surface arrival estimates for the elastic precursor are given in Supplemental Material, Sec.~VI. The first $[100]$ linear features are resolved at 8~ns, approximately 4~ns before the estimated on-axis arrival of lateral erosion. At the 12~ns probing delay, the $[100]$ measurement lies near its estimated arrival time of 12~ns, whereas the $[110]$ measurement follows its estimated arrival time of 9~ns by approximately 3~ns. Both measurements precede the estimated arrival of the elastic precursor at the rear free surface. The 12~ns radiographs therefore record linear features while parts of the diamond undergo drive-side and lateral unloading, before rear-surface release begins.

\subsection{Local structure and evolution of the simulated defect network}
\label{sec:local-structure-evolution}

Section~\ref{sec:linear-feature-timing} shows that the linear features first become visible in the $[100]$ radiographs while material near the plastic wavefront remains highly compressed, and that the 12~ns radiographs of both $[100]$ and $[110]$ targets record these features while parts of the diamond unload. Because the simulated network retains its overall organization during release, the 6~ps configuration provides a reference for the projected morphology in the 12~ns $[110]$ radiographs (Fig.~\ref{fig:fig4}). We now use the $[110]$ MD simulation to identify the local structures within this network and follow their evolution during release.

\begin{figure}
    \centering
    \includegraphics[width=1.0\textwidth]{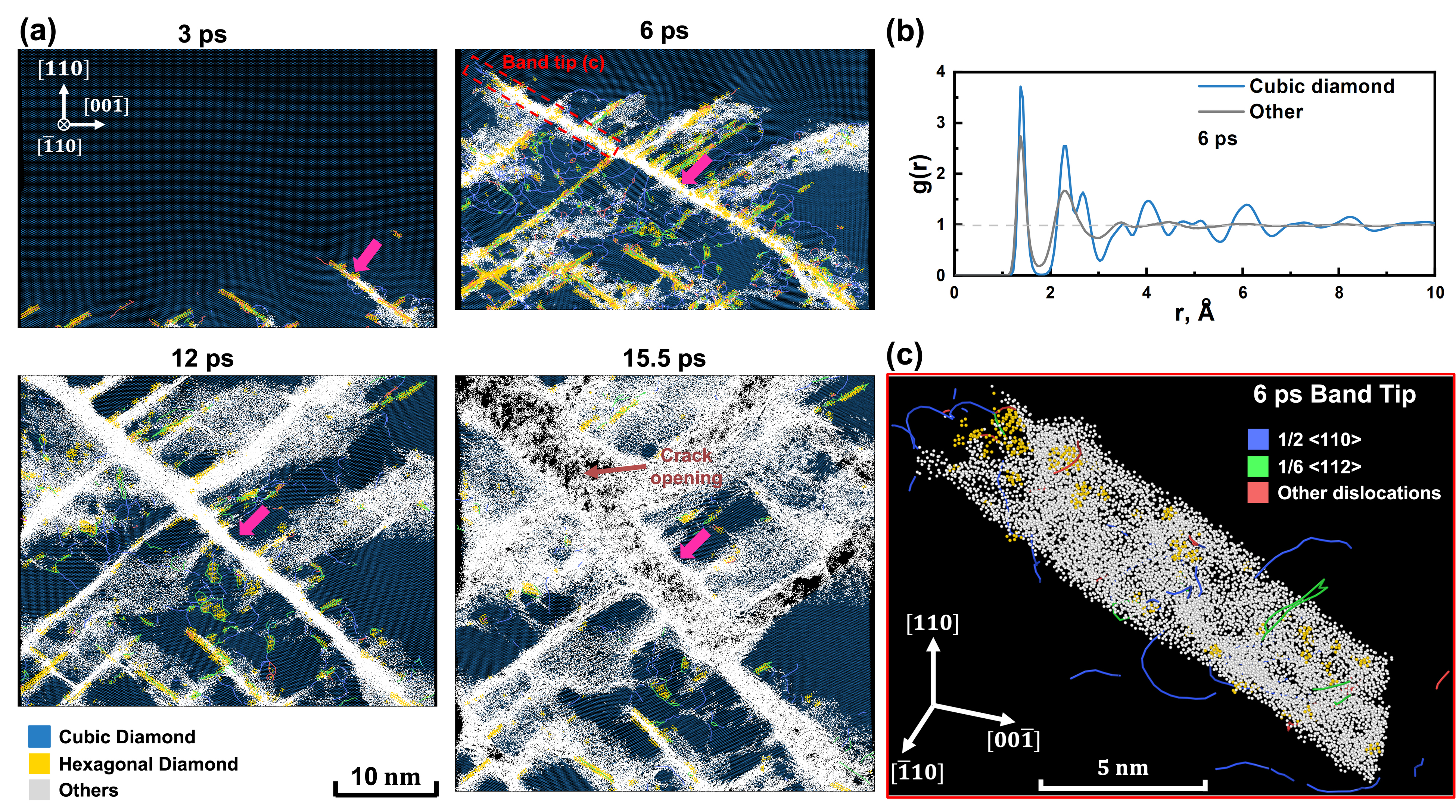}
    \caption{\textbf{Local structure and evolution of the simulated defect network during compression and release.}
    (\textbf{a}) Atomic configurations and dislocation lines in a selected ROI at 3, 6, 12, and 15.5~ps. Magenta arrows mark the same band; the red arrow marks a crack opening. Atoms classified as Cubic diamond, Hexagonal diamond, and Other are shown in blue, yellow, and gray, respectively.
    (\textbf{b}) Radial distribution functions of the cubic-diamond and Other populations within the ROI at 6~ps.
    (\textbf{c}) Enlarged view of the band tip marked in the 6~ps configuration, with cubic-diamond atoms omitted. Dislocation colors in (\textbf{a,c}) denote Burgers-vector families: blue, $\frac{1}{2}\langle110\rangle$; green, $\frac{1}{6}\langle112\rangle$; red, other.
    Scale bars are 10~nm in (\textbf{a}) and 5~nm in (\textbf{c}).}
    \label{fig:fig6}
\end{figure}

Figure~\ref{fig:fig6}(a) resolves the noncubic network behind the plastic front of Fig.~\ref{fig:fig3}(a) into narrow disordered bands connected to planes of locally hexagonal stacking. The band interiors are dominated by atoms classified as Other. At 6~ps, the radial distribution function of this population retains a pronounced first-neighbor peak but shows rapidly decaying higher-shell oscillations, whereas the cubic-diamond population within the same region of interest (ROI) retains resolved correlations over the plotted range (Fig.~\ref{fig:fig6}(b)). The retained short-range correlations and suppressed higher-shell structure support the interpretation of these regions as amorphous bands with predominantly fourfold coordination at 6~ps (Supplemental Fig.~S5(c)). The hexagonally stacked regions occupy parallel and intersecting $\{111\}$ planes connected to these bands. Their local stacking, together with their association with Shockley partial dislocations with Burgers vectors $\frac{1}{6}\langle112\rangle$, identifies these regions as stacking faults within the cubic-diamond matrix~\cite{Murri2019HexStack,Masuya2017Partial}. Together, the amorphous bands and stacking faults form an interconnected, crystallographically organized structure, referred to here as the anisotropic defect network.

The band-tip structures link amorphous-band propagation to plastic deformation in the neighboring crystal. Between 3 and 6~ps, the band marked by magenta arrows extends through the ROI while additional stacking faults develop around it and near its tip, forming a branching structure (Fig.~\ref{fig:fig6}(a)). The enlarged 6-ps view (Fig.~\ref{fig:fig6}(c)) resolves stacking faults near the band tip, together with $\frac{1}{2}\langle110\rangle$ perfect dislocations and $\frac{1}{6}\langle112\rangle$ Shockley partial dislocations in the surrounding crystal. In simulations of shocked Si, partial dislocation nucleation and propagation initiate amorphization~\cite{Zhao2016SiAmorphization}; in SiC, stacking faults accumulate and interact before amorphous bands form~\cite{Zhao2018SiCAmorphization}. The spatial association of dislocations and stacking faults with the advancing amorphous-band tip is consistent with defect-assisted amorphization, in which defect interactions generate local stresses that promote the loss of crystalline order~\cite{Zhao2016B4CDirectional,Zhao2021Directional}.

During MD simulated release, the relative contributions of amorphous bands and stacking faults to the defect network change. The band marked by magenta arrows in Fig.~\ref{fig:fig6}(a) thickens, while the surrounding stacking-fault network becomes less extensive. The atomic fractions within the ROI follow this change. The Other fraction varies only slightly between about 7 and 10~ps before increasing, while the Hexagonal diamond fraction peaks near 7.5~ps and subsequently decreases (Supplemental Fig.~S5(a)). These trends are broadly similar to those of the whole crystal (Fig.~\ref{fig:fig3}(c)); the later local maximum, compared with the whole-crystal maximum near 6~ps, is consistent with spatial variations in loading history associated with the propagating wavefronts (Fig.~\ref{fig:fig3}(a)). At 15.5~ps, the network retains its earlier spatial organization, and localized crack openings are visible within the broadened amorphous bands (red arrow in Fig.~\ref{fig:fig6}(a)), with some stacking faults still present. The amorphous bands established during compression thus become sites of local crack opening during release.

Band broadening and local crack opening during release are accompanied by changes in atomic coordination within the disordered regions. At 15.5~ps, the Other population retains short-range radial correlations but has strongly suppressed higher-shell structure relative to the cubic-diamond reference (Supplemental Fig.~S5(b)), similar to the behavior observed at 6~ps (Fig.~\ref{fig:fig6}(b)). Its coordination-number distribution, however, shifts from predominantly fourfold coordination at 6~ps to predominantly threefold coordination at 15.5~ps, with an increased fraction of twofold-coordinated atoms (Supplemental Fig.~S5(c)). The bands thus remain amorphous during release, while their coordination shifts toward lower values.

The persistence of the network's overall organization during simulated release supports using the 6~ps configuration as a structural reference for the 12~ns $[110]$ radiographs in Fig.~\ref{fig:fig4}. If the collective evolution of the network during release increases the overlap of projected density variations, individual linear features can become less distinct. This offers an interpretation of the transition from resolved linear features at 28~ns to diffuse contrast at 48~ns in the $[100]$ delay series (Fig.~\ref{fig:fig1}(b)).

Katagiri et al. interpreted the radiographic linear features in terms of stacking-fault growth driven by transonic partial-dislocation motion ~\cite{Katagiri2023Transonic}. The present atomistic analysis identifies stacking faults and dislocations as components of an interconnected network containing amorphous bands. Comparisons at matched viewing directions connect the projected morphology of this network to the observed linear contrast. During simulated release, some stacking faults persist in the crystalline regions as the amorphous bands broaden and develop local openings (Fig.~\ref{fig:fig6}(a)). Our proposed pathway provides a possible structural history for the coexistence of stacking faults and amorphous bands in recovered diamond~\cite{Li2025Direct}. In this pathway, retained stacking faults record early crystal plasticity, while band morphology and coordination also reflect subsequent unloading.

\subsection{Stress bias and tensile opening of the \texorpdfstring{$\{111\}$}{\{111\}} defect network}
\label{sec:stress-variant-selection}

\begin{figure}
    \centering
    \includegraphics[width=1.0\textwidth]{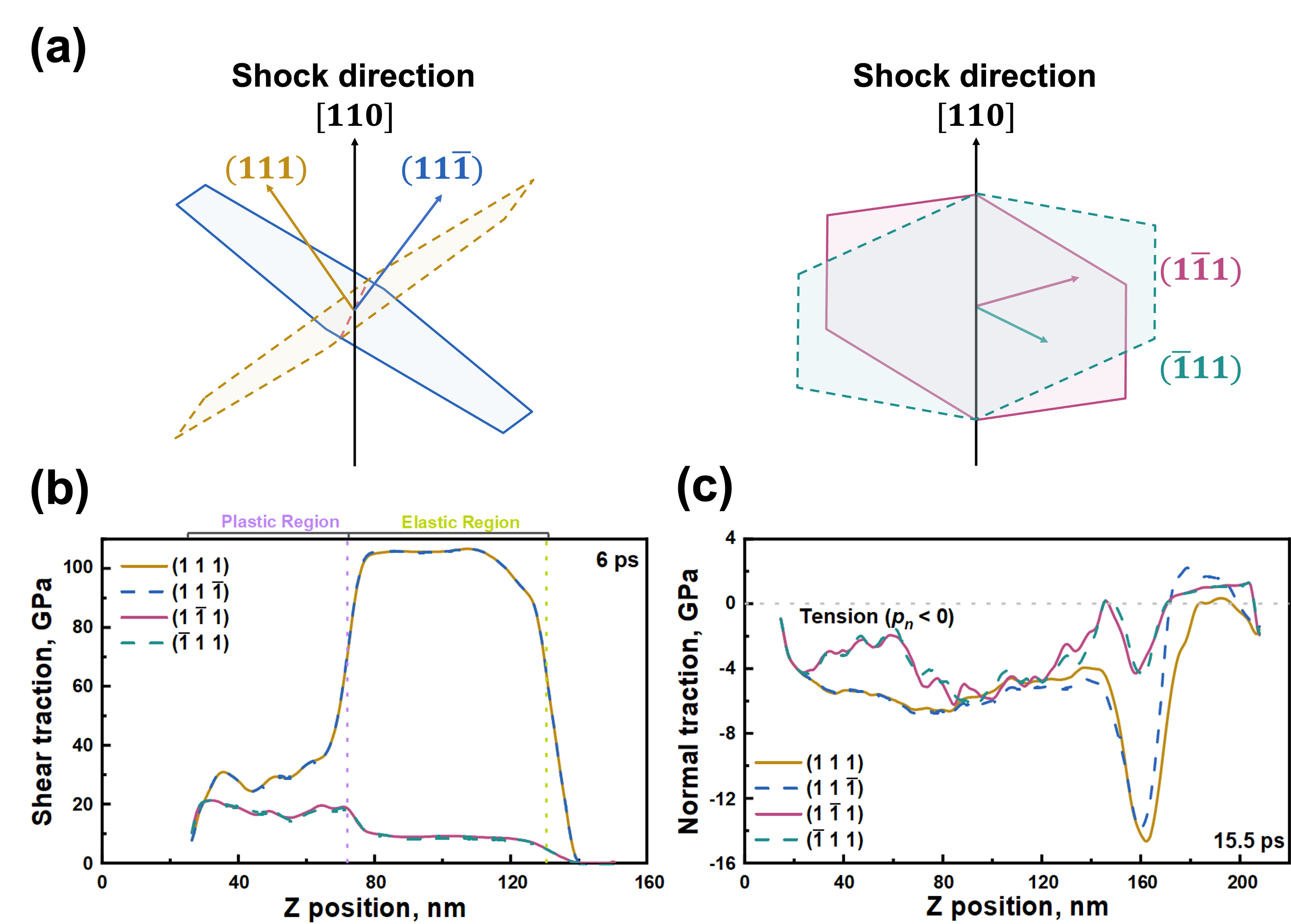}
    \caption{\textbf{Resolved tractions on the $\{111\}$ variants during compression and release.}
    (\textbf{a}) Geometry of the four reference-lattice $\{111\}$ variants relative to the $[110]$ loading direction. Colored outlines identify the planes, colored arrows their normals, and black arrows the loading direction. The red segment in the left schematic marks the intersection of $(111)$ and $(11\bar{1})$.
    (\textbf{b}) In-plane shear traction magnitudes at 6~ps. The brackets indicate the plastically and elastically compressed regions; vertical dashed lines mark the corresponding wavefronts.
    (\textbf{c}) Signed normal tractions at 15.5~ps, with compression positive and tension negative; the horizontal dotted line marks zero normal traction.}
    \label{fig:fig7}
\end{figure}

The amorphous bands and stacking faults described in Section~\ref{sec:local-structure-evolution} form preferentially along two intersecting sets of $\{111\}$ planes, with openings subsequently appearing within the network during release (Fig.~\ref{fig:fig6}(a)). This sequence raises two related questions about why the network develops along these particular plane variants and how subsequent unloading promotes opening within it. We compare the in-plane shear-traction magnitude $\tau$ during compression and the signed normal traction $p_n$ during release on the four reference-lattice $\{111\}$ variants. These quantities are calculated from the averaged virial pressure tensor (Supplemental Material, Sec.~V). The four symmetry-equivalent $\{111\}$ variants in the undeformed cubic crystal are divided into two geometrically distinct pairs relative to the $[110]$ shock direction. As shown in Fig.~\ref{fig:fig7}(a), the $[110]$ shock direction lies within both the $(1\bar{1}1)$ and $(\bar{1}11)$ planes, whereas the $(111)$ and $(11\bar{1})$ planes are inclined to it. In the reference view along $[\bar{1}10]$, the $(111)$ and $(11\bar{1})$ planes are edge-on, with oppositely inclined traces consistent with the two families in the MD structure maps.

The geometrical distinction between $\{111\}$ variants becomes mechanically important because longitudinal compression substantially exceeds transverse compression between the elastic and plastic fronts (Fig.~\ref{fig:fig3}(b)). At 6~ps, the $(111)/(11\bar{1})$ pair carries in-plane shear tractions of about $105~\mathrm{GPa}$ in the elastically compressed region, compared with about $10~\mathrm{GPa}$ on the other two variants (Fig.~\ref{fig:fig7}(b)). Behind the plastic front, increased transverse compression reduces the stress contrast between the longitudinal and transverse directions, thereby lowering the shear traction on this pair and reducing its separation from the other two variants. The shear bias is therefore strongest in the elastically compressed material ahead of the plastic front and persists at a lower magnitude in the plastically compressed region where the defect network develops. The stacking faults and amorphous bands form preferentially on the $(111)$ and $(11\bar{1})$ variants, which carry the larger in-plane shear tractions. Together with the dislocation activity associated with the amorphous bands (Fig.~\ref{fig:fig6}(c)), this correspondence supports a mechanical contribution to the plane-variant preference.

Release subsequently acts on a crystal containing a network of amorphous bands and stacking faults formed preferentially on the $(111)$ and $(11\bar{1})$ planes during compression. At 15.5~ps (shown in Fig.~\ref{fig:fig7}(c)), the normal tractions on all four variants are negative over much of the sampled material, indicating tension under the compression-positive convention used in MD simulation. While tensile loading is not restricted to the dominant network variants, the $(111)/(11\bar{1})$ pair exhibits the deeper tensile minimum. The observed crack openings occur within amorphous bands that were already present during compression (Fig.~\ref{fig:fig6}(a)). Their location within the pre-existing network, together with the tensile normal loading, supports the role of the amorphous bands as paths for local separation during release. Shear stress thus biases the formation of the defect network during compression, and normal tension subsequently favors opening within its amorphous bands.

\subsection{Three-dimensional band morphology from viewing-direction dependence}
\label{sec:viewing-direction-geometry}

The structural analysis in Section~\ref{sec:local-structure-evolution} identifies amorphous bands connected to stacking faults on the $(111)$ and $(11\bar{1})$ variants, and the traction analysis in Section~\ref{sec:stress-variant-selection} relates this plane preference to shear stress. At $\phi=0^\circ$, both variants are viewed edge-on. Sheets with different in-plane shapes on a given plane can therefore share the same projected orientation in this view. We use the rotation series in Figs.~\ref{fig:fig2} and \ref{fig:fig4} to examine their in-plane morphology through changes in projected feature angles and visibility in experiment and MD. The rotation angle $\phi$ specifies the viewing direction about the $[110]$ shock axis, with $\phi=0^\circ$ corresponding to a view along $[\bar{1}10]$. The projected feature angle $\beta$ is measured from the detector horizontal.

\begin{figure}
    \centering
    \includegraphics[width=\textwidth]{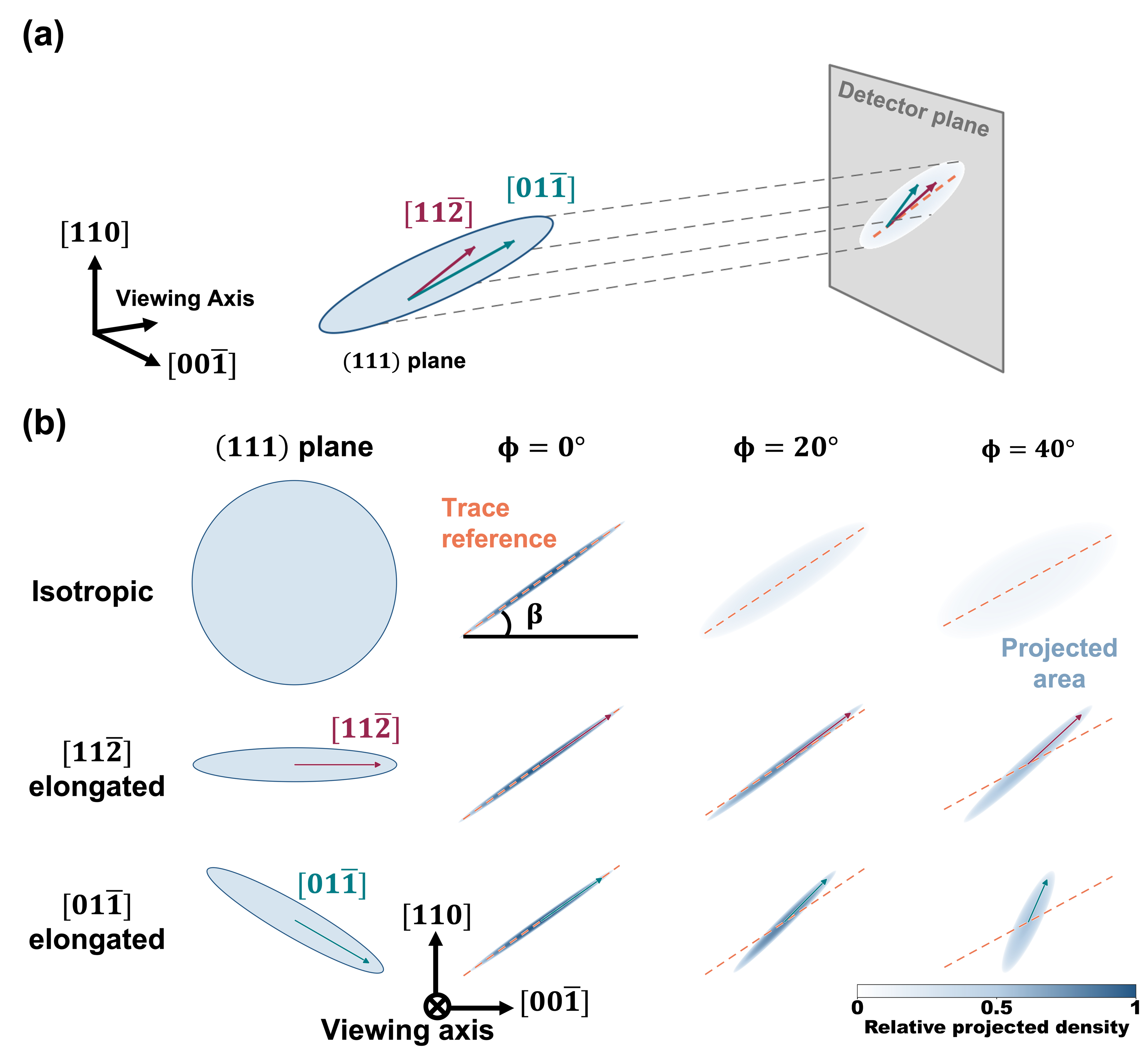}
    \caption{\textbf{Projection geometry and viewing-direction dependence of finite sheets.}
    (\textbf{a}) Geometry of a finite $(111)$ sheet and its detector projection.
    (\textbf{b}) In-plane shapes and model projections at $\phi=0^\circ$, $20^\circ$, and $40^\circ$ for rotation about $[110]$. At $\phi=0^\circ$, the viewing direction is parallel to $[\bar{1}10]$. Rows show an in-plane isotropic circular sheet and elliptical sheets elongated along $[11\bar{2}]$ and $[01\bar{1}]$, from top to bottom. The projected major-axis angle $\beta$ is measured from the detector horizontal. Orange dashed lines mark the $(111)$ plane trace; colored arrows indicate the in-plane reference directions and their projections. Blue shading represents the model density integrated along the line of sight, normalized within each row to its peak at $\phi=0^\circ$ using a common linear color scale.}
    \label{fig:fig8}
\end{figure}

Figure~\ref{fig:fig8}(a) shows the projection geometry of a finite $(111)$ sheet, used here as a geometric representation of the band morphology resolved in Fig.~\ref{fig:fig6}(c). Figure~\ref{fig:fig8}(b) compares three uniform-density sheets with different in-plane shapes and their projections at $\phi=0^\circ$, $20^\circ$, and $40^\circ$. The circular sheet represents an in-plane isotropic shape. The two elliptical sheets are elongated along the in-plane directions $[11\bar{2}]$ and $[01\bar{1}]$ of $(111)$, which belong to the $\langle112\rangle$ and $\langle110\rangle$ families, respectively. Their projections are line-of-sight integrals of the model density and provide geometric references for projected orientation and broadening. The model does not include phase-contrast image formation (Supplemental Material, Sec.~IX).

At $\phi=0^\circ$, the $(111)$ plane is viewed edge-on, and all three shapes produce narrow projections with the same major-axis orientation. For the circular sheet, the projected major axis follows the plane trace, defined by the intersection of the sheet plane with the detector plane. Its absolute projected angle $|\beta|$ decreases as $\phi$ increases. For the elongated sheets shown, the projected major axes remain close to the projections of their in-plane long axes. Their $|\beta|$ values increase with $\phi$, moderately for elongation along $[11\bar{2}]$ and more rapidly for elongation along $[01\bar{1}]$. The sign and magnitude of the change in $|\beta|$ with increasing $\phi$ therefore distinguish these model morphologies.

\begin{figure}
    \centering
    \includegraphics[width=\textwidth]{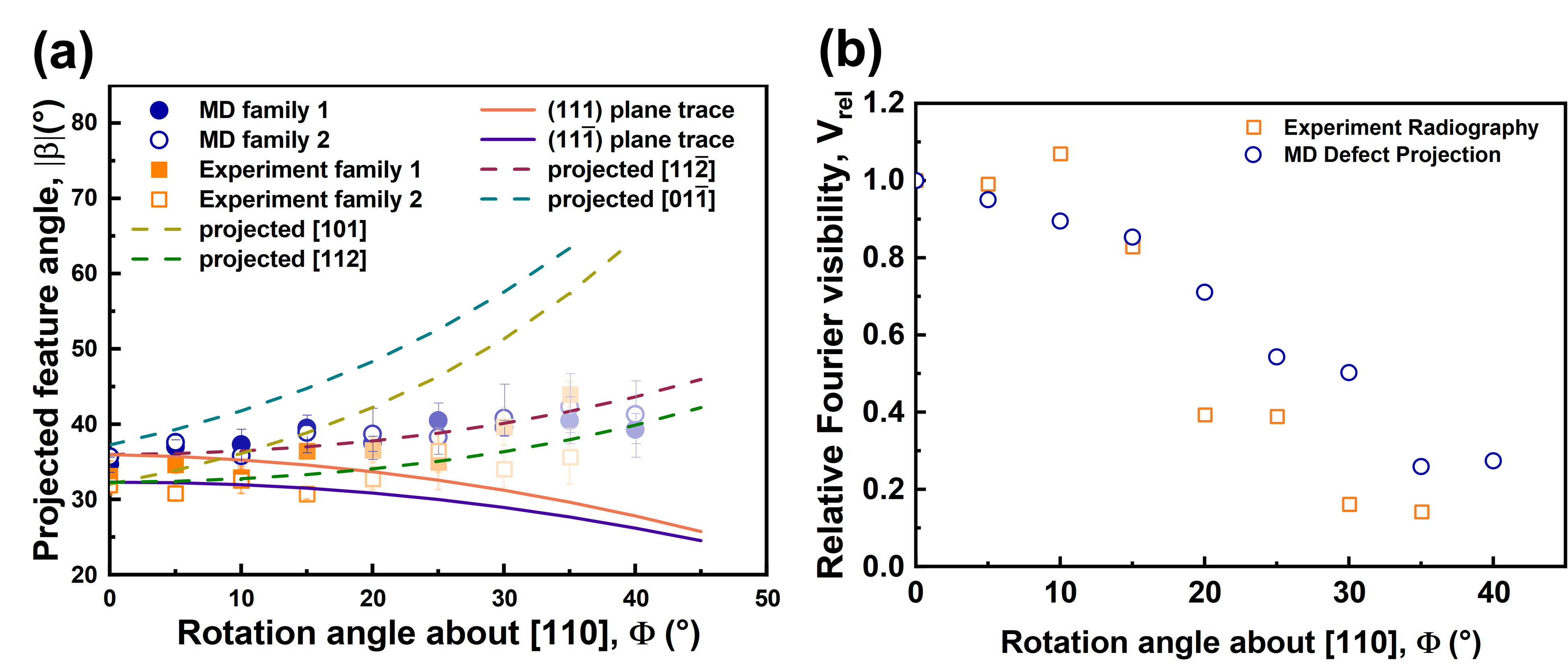}
    \caption{\textbf{Projected feature angles and Fourier visibility during rotation about $[110]$.}
    (\textbf{a}) Absolute projected feature angles, $|\beta|$, for experiment and MD as a function of the crystal rotation angle $\phi$. The angle $\beta$ is measured from the detector horizontal. Orange squares and blue circles denote experimental radiographs and MD projections, respectively. Filled and open symbols denote Families~1 and~2, corresponding to the $(11\bar{1})$ and $(111)$ variants, respectively. Error bars show the standard deviation of measured segment angles for experiment and the selected Fourier peak-center ranges for MD. Solid and dashed curves show deformation-corrected plane-trace and projected in-plane direction references, respectively, calculated using local deformation gradients from the 6~ps MD configuration. Symbol opacity reflects the corresponding dataset visibility in (\textbf{b}).
    (\textbf{b}) Relative Fourier visibility, $V_{\mathrm{rel}}$, calculated using angular windows centered on directions perpendicular to the ideal-crystal projections of $[112]$ and $[11\bar{2}]$. Values are averaged over the two feature families and normalized to each dataset's value at $\phi=0^\circ$.}
    \label{fig:fig9}
\end{figure}

Figure~\ref{fig:fig9}(a) compares the measured absolute projected feature angles, $|\beta|$, with plane-trace and projected in-plane direction references for both the $(111)$ and $(11\bar{1})$ variants. Experimental angles for $\phi=0^\circ$--$35^\circ$ are obtained from line segments in Fourier-filtered radiographs. MD angles are obtained from peak centers in radially whitened Fourier angular-power distributions, with the Fourier-normal directions converted to real-space feature orientations. Radial whitening reduces the influence of radial variations in Fourier power on the angular analysis. The extraction procedures are described in Supplemental Material, Secs.~I and~IV, with a representative MD extraction shown in Supplemental Fig.~S2. 

Families~1 and~2, labeled in the $\phi=0^\circ$ radiograph in Fig.~\ref{fig:fig2}, are associated with the $(11\bar{1})$ and $(111)$ variants, respectively, based on the signs of their projected angles $\beta$ at $\phi=0^\circ$. The in-plane reference directions are $[112]$ and $[101]$ for $(11\bar{1})$, and $[11\bar{2}]$ and $[01\bar{1}]$ for $(111)$. In ideal cubic geometry, the paired references for the two variants have equal $|\beta|$ at each $\phi$, because their detector projections are related by mirror symmetry about the vertical axis. Figure~\ref{fig:fig9}(a) uses reference curves calculated with local deformation gradients from the plastically compressed region of the 6~ps MD configuration (Supplemental Material, Sec.~VIII). Local shear and lattice rotation can break the ideal mirror symmetry. The resulting curves therefore provide separate geometric references for assessing the in-plane elongation of each family in the deformed MD lattice.

Because the ideal sheet models shown in Fig.~\ref{fig:fig8} share the same projected orientation at $\phi=0^\circ$, we compare how $|\beta|$ changes with increasing $\phi$ relative to this reference view. Both MD families show a moderate overall increase in $|\beta|$ and remain close to one another. Both experimental families also increase overall, with a larger change for Family~1 than for Family~2. These angular trends are more consistent with the moderately increasing $\langle112\rangle$ references than with the decreasing plane-trace references or the more rapidly increasing $[101]$ and $[01\bar{1}]$ references. Within the tested models, the trends favor finite sheets elongated near $[112]$ on $(11\bar{1})$ and near $[11\bar{2}]$ on $(111)$, consistent with the defect-band morphology in Fig.~\ref{fig:fig6}(c).

Figure~\ref{fig:fig8}(b) also shows how the projected widths and peak densities change with viewing direction. As the viewing direction moves away from the edge-on geometry, the projected shapes broaden and their peak projected densities decrease, while the integrated projected density remains constant (Supplemental Material, Sec.~IX). Within a complex 3D network, the broadened contributions of neighboring bands and connecting segments can overlap, making individual features less distinct. Figure~\ref{fig:fig4}(c) shows projections of the same 6~ps atomic configuration from different viewing directions. The decreasing visibility of individual features in this series therefore demonstrates how viewing geometry changes the appearance of an existing network.

Figure~\ref{fig:fig9}(b) quantifies the oblique directional contrast in the Fourier spectra of Fig.~\ref{fig:fig4}(b,d) and Supplemental Fig.~S4 using the relative Fourier visibility, $V_{\mathrm{rel}}$. At each $\phi$, the angular windows are centered on directions perpendicular to the ideal-crystal projections of $[112]$ and $[11\bar{2}]$. These references correspond to the candidate $\langle112\rangle$ elongation directions favored by the angle comparison. The values are averaged over the two families and normalized to each dataset's value at $\phi=0^\circ$ (Supplemental Material, Sec.~X). Both experiment and MD show an overall decrease in $V_{\mathrm{rel}}$ as $\phi$ increases, accompanying the decreasing visibility of individual linear features in real space. At $\phi=20^\circ$, $V_{\mathrm{rel}}$ is approximately 0.4 for experiment and 0.7 for MD, showing a larger reduction in experimental directional contrast at this rotation.

Supplemental Fig.~S3 shows viewing-direction sensitivity in the additional $[100]$ rotation series. At a pump-probe delay of 12~ns, the intersecting linear features are clearly resolved near the reference $[011]$ view and become less distinct at larger rotations about $[100]$. The rotation-dependent visibility under both loading directions, together with the $[110]$ projected-angle trends and the fixed-configuration MD projections, supports a finite-sheet description of the structures producing the linear contrast and a geometric contribution to their decreasing visibility.

\section{Conclusion}
Radiographs acquired at multiple pump-probe delays resolve the appearance and extension of linear features behind the plastic wavefront in diamond shocked along $[100]$. Hydrodynamic calculations show that material near the plastic wavefront remains highly compressed when these features first become visible, while regions farther behind the front unload. Geometric analysis of the $[110]$ rotation series favors finite $\{111\}$ sheets with preferential elongation along in-plane directions close to $\langle112\rangle$. In the $[110]$ MD simulation, amorphous bands and stacking faults form an interconnected network preferentially occupying the $\{111\}$ variants with larger in-plane shear traction. The association of dislocations with band tips is consistent with defect-assisted amorphization. Projections of this network show a loss of feature visibility with rotation similar to experiment. During simulated release, tensile loading opens local cracks within the amorphous bands.

Together, our results support a pathway in which crystal plasticity and localized amorphization create an anisotropic defect network during compression, with local cracks subsequently opening within its amorphous bands under tensile loading during release. In this pathway, the same defect network connects deformation under compression to failure during release, with its crystallographic organization linking the loading geometry to preferred damage sites. By characterizing the morphology and connectivity of this network and its evolution during release, the present results provide microstructural constraints for models of dynamic strength and failure in strongly covalent solids. Shock-induced amorphous bands reported in boron carbide and silicon~\cite{Chen2003Shock,Zhao2016SiAmorphization} motivate examining the connection between compression and release mechanisms beyond diamond.

\begin{acknowledgments}
Z.X. thanks Meizhong Lyu, Brinthan Kanesalingam and Yifan Wang for helpful discussions
and feedback on this work.

This work was supported by the Air Force Office of Scientific Research (AFOSR) under Contract No.~FA9550-23-1-0347. This work used Bridges-2 at Pittsburgh Supercomputing Center through allocation MAT250094 from the Advanced Cyberinfrastructure Coordination Ecosystem: Services \& Support (ACCESS) program, which is supported by U.S. National Science Foundation grants \#2138259, \#2138286, \#2138307, \#2137603, and \#2138296~\cite{Brown2021Bridges2,Boerner2023ACCESS}.

The experiments were performed at BL3 of SACLA with the approval of the Japan Synchrotron Radiation Research Institute (proposal nos.~2021A8004, 2021B8067, and 2024A8004). The high-power nanosecond laser at SACLA was installed by the Institute of Laser Engineering of the The University of Osaka in cooperation with Hamamatsu Photonics. The installation of diffractive optical elements (DOEs) to improve the smoothness of the drive laser pattern was supported by the SACLA Basic Development Program. This work was partially supported through the use of research equipment shared under the MEXT Project for Promoting Public Utilization of Advanced Research Infrastructure (Program for advanced research equipment platforms), Grant Nos.~JPMXS0450300321 and JPMXS0450300324. This work was also supported by KAKENHI (Grant Nos. 21K03499 and 24K06988) from the Japan Society for the Promotion of Science (JSPS).
\end{acknowledgments}

\section*{Author Contributions}
Z.X., K.K., and L.E.D.-M. conceptualized the study. Z.X., K.K., and T.P. developed the methodology. Z.X., K.K., T.P., A.H., D.P.L., S.E.P., Y.I., K.M., H.N., T.T., M.Y., T.Y., and N.O. carried out the investigation. T.P., Y.I., K.M., H.N., T.T., M.Y., T.Y., and N.O. contributed to curation of the primary data, and Z.X. curated the combined dataset used in this study. Z.X. performed the formal analyses reported in this work, validated the analysis, prepared the visualizations, and developed the analysis software. A.H., D.P.L., and S.E.P. contributed to the analysis and interpretation of the results. W.L. provided guidance on the molecular-dynamics simulations and contributed to validation of the simulation. T.P., W.L., and L.E.D.-M. provided supervision. N.O. and L.E.D.-M. provided resources and acquired funding. K.K., N.O., and L.E.D.-M. administered the project. Z.X. wrote the original draft. K.K., T.P., W.L., A.H., D.P.L., S.E.P., N.O., and L.E.D.-M. reviewed and edited the manuscript.

\section*{Competing Interests}

The authors declare that they have no competing interests.

\section*{Data Availability}

The experimental radiographs supporting the findings of this study will be deposited in Zenodo upon publication. The raw molecular-dynamics trajectories are not publicly archived because of their large file size and are available from the corresponding author upon reasonable request.

\bibliography{references}

\end{document}

\title{Supplemental Material for\texorpdfstring{\\}{ }
Emergence of Anisotropic Defect Networks in Shocked Diamond}

\author{Zipeng Xu}
\email[Co-corresponding author: ]{zipengx@stanford.edu}
\affiliation{Department of Materials Science and Engineering, Stanford University, Stanford, California 94305, USA}
\affiliation{SLAC National Accelerator Laboratory, Menlo Park, California 94025, USA}
\affiliation{PULSE Institute, Stanford University, Stanford, California 94305, USA}

\author{Kento Katagiri}
\affiliation{Department of Materials Science and Engineering, Stanford University, Stanford, California 94305, USA}
\affiliation{SLAC National Accelerator Laboratory, Menlo Park, California 94025, USA}
\affiliation{PULSE Institute, Stanford University, Stanford, California 94305, USA}
\affiliation{Graduate School of Engineering, The University of Osaka, Osaka 565-0871, Japan}
\affiliation{Institute of Laser Engineering, The University of Osaka, Osaka 565-0871, Japan}

\author{Tatiana Pikuz}
\affiliation{Institute for Open and Transdisciplinary Research Initiatives, The University of Osaka, Suita, Osaka 565-0871, Japan}
\affiliation{Institute for Laser Technology, Suita, Osaka 565-0871, Japan}

\author{Wanghui Li}
\affiliation{School of Civil Engineering and Transportation, South China University of Technology, Guangzhou 510640, China}
\affiliation{Institute of High Performance Computing (IHPC), Agency for Science, Technology and Research (A*STAR), Singapore 138632, Republic of Singapore}

\author{Anirudh Hari}
\affiliation{Department of Materials Science and Engineering, Stanford University, Stanford, California 94305, USA}
\affiliation{SLAC National Accelerator Laboratory, Menlo Park, California 94025, USA}
\affiliation{PULSE Institute, Stanford University, Stanford, California 94305, USA}

\author{Dorian P. Luccioni}
\affiliation{Department of Materials Science and Engineering, Stanford University, Stanford, California 94305, USA}
\affiliation{SLAC National Accelerator Laboratory, Menlo Park, California 94025, USA}
\affiliation{PULSE Institute, Stanford University, Stanford, California 94305, USA}

\author{Sophie E. Parsons}
\affiliation{Department of Materials Science and Engineering, Stanford University, Stanford, California 94305, USA}
\affiliation{SLAC National Accelerator Laboratory, Menlo Park, California 94025, USA}
\affiliation{PULSE Institute, Stanford University, Stanford, California 94305, USA}

\author{Yuichi Inubushi}
\affiliation{Japan Synchrotron Radiation Research Institute, Sayo 679-5198, Japan}
\affiliation{RIKEN SPring-8 Center, Sayo 679-5148, Japan}

\author{Kohei Miyanishi}
\affiliation{RIKEN SPring-8 Center, Sayo 679-5148, Japan}

\author{Hirotaka Nakamura}
\affiliation{Graduate School of Engineering, The University of Osaka, Osaka 565-0871, Japan}

\author{Tadashi Togashi}
\affiliation{Japan Synchrotron Radiation Research Institute, Sayo 679-5198, Japan}
\affiliation{RIKEN SPring-8 Center, Sayo 679-5148, Japan}

\author{Makina Yabashi}
\affiliation{Japan Synchrotron Radiation Research Institute, Sayo 679-5198, Japan}
\affiliation{RIKEN SPring-8 Center, Sayo 679-5148, Japan}

\author{Toshinori Yabuuchi}
\affiliation{Japan Synchrotron Radiation Research Institute, Sayo 679-5198, Japan}
\affiliation{RIKEN SPring-8 Center, Sayo 679-5148, Japan}

\author{Norimasa Ozaki}
\affiliation{Graduate School of Engineering, The University of Osaka, Osaka 565-0871, Japan}
\affiliation{Institute of Laser Engineering, The University of Osaka, Osaka 565-0871, Japan}

\author{Leora E. Dresselhaus-Marais}
\email[Co-corresponding author: ]{leoradm@stanford.edu}
\affiliation{Department of Materials Science and Engineering, Stanford University, Stanford, California 94305, USA}
\affiliation{SLAC National Accelerator Laboratory, Menlo Park, California 94025, USA}
\affiliation{PULSE Institute, Stanford University, Stanford, California 94305, USA}

\maketitle

\section{Linear feature extraction - Experimental radiography}
\label{sec:linear-feature-angle-extraction-exp}

\begin{figure}[!hbp]
    \centering
    \includegraphics[width=1.0\textwidth]{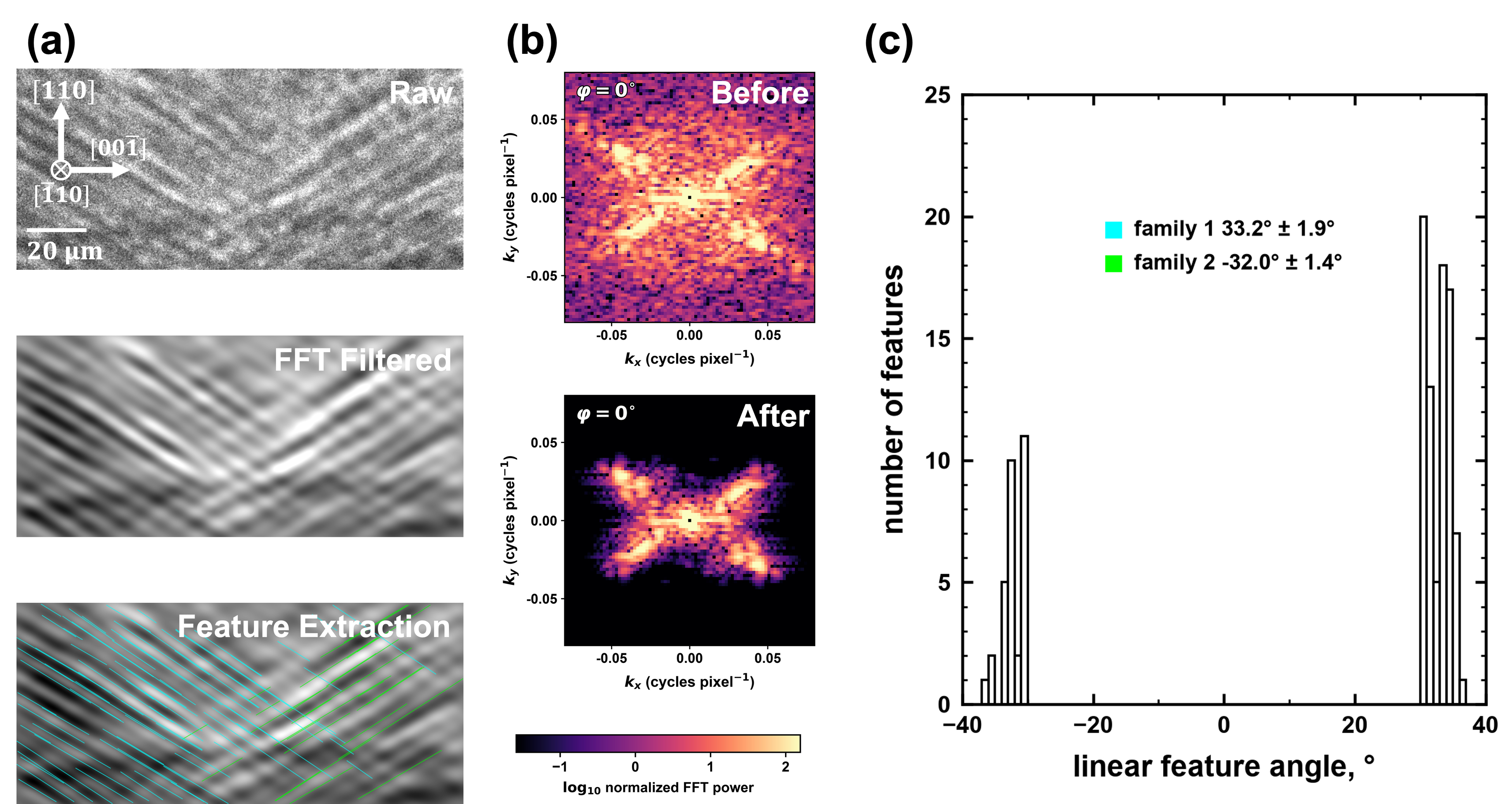}
    \caption{
    \textbf{Fourier filtering and linear feature angle extraction from experimental radiographs.}
    A representative analysis is shown for a radiograph for [110] single-crystal diamond.
    (\textbf{a}) Cropped radiograph before filtering (top), the corresponding inverse-FFT reconstruction (middle), and retained line segments overlaid on the reconstructed image  (bottom). Cyan and green segments correspond to Family~1 and Family~2, respectively. The 20~$\mu$m scale bar applies to all three images.
    (\textbf{b}) Corresponding Fourier power spectra before (top) and after (bottom) intensity threshold filtering, displayed on a logarithmic color scale.
    (\textbf{c}) Distributions of the signed segment angle $\beta$, measured from the detector horizontal, with clockwise rotation taken as positive. The values shown for each family are the mean angle and standard deviation.
    }
    \label{fig:linear_feature_Exp}
\end{figure}

Following the general strategy used in Ref.~\cite{SM-Katagiri2023Transonic}, radiographic linear-feature orientations were extracted by Fourier filtering followed by Hough transform-based line segment detection. Regions of interest behind the plastic wavefront were cropped in ImageJ and processed using a two-dimensional fast Fourier transform (FFT). An intensity threshold was manually adjusted on the displayed Fourier power spectrum to select the conjugate intensity lobes associated with the two feature families. The threshold-defined selection was used for Fourier filtering, and the filtered image was reconstructed by inverse FFT. A representative analysis is shown in Fig.~\ref{fig:linear_feature_Exp}.

The reconstructed images were linearly normalized, converted to 8-bit intensity, and smoothed with a $3\times3$ Gaussian kernel with a standard deviation of 0.5 pixel. Edges were detected using the Canny algorithm with lower and upper thresholds of 15 and 50, respectively. Line segments were identified using the probabilistic Hough transform implemented in OpenCV as \texttt{HoughLinesP}, with distance and angular resolutions of 1 pixel and $1^\circ$, respectively. The Hough vote threshold and minimum segment length were 25 and 20 pixels for $\phi=0^\circ$--$10^\circ$, and 20 and 15 pixels for $\phi=15^\circ$--$35^\circ$. At $\phi=40^\circ$, no reliable Hough-based line-segment angle was reported, but the radiograph was retained for analysis of its Fourier power spectrum.

The signed segment angle $\beta$ was measured relative to the detector horizontal. Segments satisfying $30^\circ<|\beta|<60^\circ$ were retained. This interval was selected based on the expected inclinations of $\{111\}$-related features in the reference viewing geometry, with allowance for deformation-induced angular deviations. The retained segments were grouped by angle using DBSCAN with $\epsilon=3^\circ$ and \texttt{min\_samples}=2. The two clusters containing the most segments were assigned to the two crossing feature families. For each family, the reported angle was calculated as the mean of the retained segment angles, with error bars representing one population standard deviation of the segment-angle distribution.

\clearpage
\section{MD initialization and time integration}
\label{sec:md-simulation-details}

The simulation geometry, interatomic potential, and loading history are specified in Sec.~II\,B of the main text. The initial crystal was generated using Atomsk~\cite{SM-hirel2015atomsk} and energy-minimized with box relaxation to zero pressure. Equilibration was performed for 30~ps in the NPT ensemble at 300~K and zero pressure, using periodic boundary conditions in all three directions and a time step of 1~fs.

For shock loading, the boundary condition along $z$ was changed from periodic to shrink-wrapped nonperiodic. The bulk crystal was evolved in the NVE ensemble with a time step of 0.5~fs throughout the driven and release stages.

\section{MD defect projection}
\label{sec:md-defect-projection}

Two-dimensional defect fraction maps were generated from the MD configuration at 6~ps (step 12000). Atomic environments were classified using polyhedral template matching (PTM) in OVITO. Atoms assigned to Structure Types 0 and 7, corresponding to Other and Hexagonal diamond, respectively, were treated as defect atoms. Here, ``defect atoms'' denotes a projection-analysis category combining these local environments; it does not imply that all selected atoms represent the same microscopic defect type.

At $\phi=0^\circ$, the projection direction was parallel to $[\bar{1}10]$, the detector horizontal was parallel to $[00\bar{1}]$, and the detector vertical and rotation axis were parallel to $[110]$. The atomic configuration was rotated about the fixed $[110]$ axis from $\phi=0^\circ$ to $40^\circ$ in $5^\circ$ increments. All atoms in each rotated configuration were projected onto a common detector grid with $5\times5~\text{\AA}^2$ pixels.

For each occupied detector pixel, the projected defect fraction was calculated as the number of defect atoms divided by the total number of atoms projected into that pixel. The fraction maps were then smoothed with a Gaussian kernel with a full width at half maximum of $10~\text{\AA}$. The resulting maps were used to compare the projected morphology and viewing-direction dependence of the simulated defect network with the experimental radiographs.

\section{Linear feature extraction - MD defect projection}
\label{sec:linear-feature-angle-extraction-md}

Linear-feature orientations were extracted from the two-dimensional Fourier power spectra of the projected defect fraction maps. For the MD projections, $y$ and $z$ denote the horizontal and vertical image coordinates, respectively. The same region of interest, $87.5 \le y \le 757.5$~\AA{} and $322.5 \le z \le 702.5$~\AA{}, was used for every rotation angle. The Fourier angular power distribution was obtained by radial whitening. The Fourier-lobe azimuth $\psi_{\mathrm{FFT}}$ corresponds to the normal direction of the real-space linear features. It was converted to the signed real-space feature angle $\beta$, measured from the detector horizontal.

\begin{figure}[!hbp]
    \centering
    \includegraphics[width=1.0\textwidth]
    {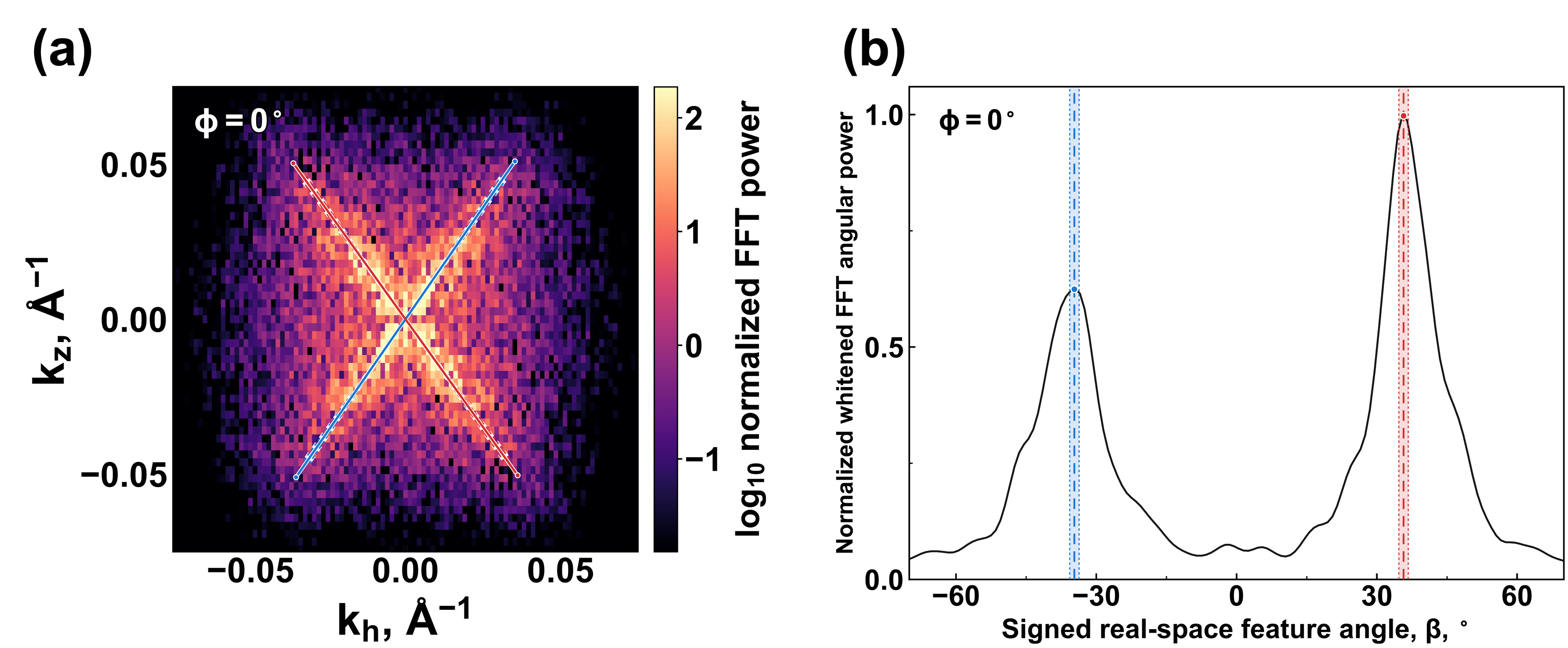}
    \caption{
    \textbf{Fourier-space extraction of linear-feature angles from an MD defect projection.}
    A representative analysis is shown for the $[110]$ MD defect projection at $\phi=0^\circ$.
    (\textbf{a}) Unwhitened two-dimensional Fourier power spectrum of the MD defect projection, normalized by its median power and displayed on a logarithmic scale. The solid blue and red diameters indicate the Fourier-normal directions corresponding to the selected real-space feature angles.
    (\textbf{b}) Radially whitened angular-power distribution averaged over the MD defect projection, normalized by its maximum, and plotted against the signed real-space feature angle $\beta$ relative to the detector horizontal. The blue and red dashed lines and symbols mark the selected centers at $\beta=-34.7^\circ$ and $+35.7^\circ$, respectively. The shaded intervals and dotted boundaries show the corresponding
error bar range.}
    \label{fig:linear_feature_MD}
\end{figure}

The two signed feature branches were identified by jointly inspecting the angular-power distribution and the corresponding directional lobes in the two-dimensional spectrum. For a sharp, well-resolved peak, the representative angle $\beta_0$ was taken at the dominant local maximum. For broadened, split, or flat-topped peaks, a representative center was selected while following the corresponding branch across the rotation series. The lower and upper angular bounds, $\beta_{\mathrm{low}}$ and $\beta_{\mathrm{high}}$, were selected to span the candidate peak-center positions associated with the same Fourier lobe, accounting for nearby maxima, shoulders, and flat peak tops. A representative spectrum, angular-power distribution, and selected angular ranges are shown in Fig.~\ref{fig:linear_feature_MD}.

\section{Calculation of plane-resolved tractions}
\label{sec:plane-traction-analysis}

The plane-resolved tractions in Fig.~7 were calculated
by projecting the spatially and temporally averaged virial stress
tensor $\boldsymbol{\sigma}_{\mathrm{vir}}$ onto the four
reference-lattice $\{111\}$ variants: $(111)$, $(11\bar{1})$,
$(1\bar{1}1)$, and $(\bar{1}11)$. Using the compression-positive tensor
$\mathbf{P}=-\boldsymbol{\sigma}_{\mathrm{vir}}$, the normal traction
$p_n$ and in-plane shear-traction magnitude $\tau$ were calculated as
\begin{align}
    p_n
    &= \hat{\mathbf n}^{T}\mathbf P\hat{\mathbf n},
    \label{eq:normal-traction}
    \\
    \tau
    &= \left\|
    \mathbf P\hat{\mathbf n}-p_n\hat{\mathbf n}
    \right\|,
    \label{eq:shear-traction}
\end{align}
where $\hat{\mathbf n}$ is the unit plane normal defined by the initial
crystal orientation and expressed in the simulation coordinate system.
Positive and negative $p_n$ indicate compressive and tensile normal
loading, respectively.

\section{Lateral-release and rear-surface-release timing estimates}
\label{sec:timing-estimate}

Lateral release originates at the edge of the finite laser-driven region. Its erosion of the advancing plastic wavefront is estimated using the construction of Swift and Kraus \cite{SM-Swift2008PlasticAblators}. The angles $\gamma_1$ and $\gamma_2$ describe shock erosion in the PP and diamond, respectively, and are measured relative to the shock direction. All times are referenced to drive-laser arrival at the front surface of the PP ablator. Shock entry into diamond is taken to occur at approximately 4~ns for both loading orientations. The calculations used a drive-spot diameter of $260~\mu\mathrm{m}$, a PP thickness of $h_{\mathrm{PP}}=50~\mu\mathrm{m}$, and a diamond thickness of $h_d=250~\mu\mathrm{m}$. Laser-off release and interface release are evaluated in a 1D HELIOS simulation. 

Following Swift and Kraus~\cite{SM-Swift2008PlasticAblators}, the shock-erosion angle $\gamma$, measured relative to the shock direction, is given by
\begin{equation}
\tan\gamma=
\frac{\sqrt{c^2-(U_s-u_p)^2}}{U_s},
\label{eq:shock-erosion-angle}
\end{equation}
where $c$ is the sound speed in the shocked material, $U_s$ is the shock speed, and $u_p$ is the particle velocity behind the shock.

During Stage~1, the shock propagates through the PP ablator, while lateral release from the edge of the driven region narrows the nominal peak-compression region. An erosion angle of $\gamma_1=45^\circ$ was adopted for the PP layer in the geometric construction. Over the $50~\mu\mathrm{m}$ ablator thickness, this construction reduces the driven half-width from $130~\mu\mathrm{m}$ at the front surface to $R_0=80~\mu\mathrm{m}$ at the PP--diamond interface.

Stage~2 begins when the shock enters the diamond. The diamond erosion angle $\gamma_2$ was estimated using the elastic-wave speed $c_e$ and the plastic-wave speed $D_p$. Neglecting the motion of the compressed material in this geometric approximation gives
\begin{equation}
\tan\gamma_2=
\frac{\sqrt{c_e^2-D_p^2}}{D_p}.
\label{eq:diamond-erosion-angle}
\end{equation}
The interface half-width $R_0$ and angle $\gamma_2$ determine the axial depth $d_1$ at which the modeled lateral-release boundaries meet on the centerline. The corresponding time $t_{\mathrm{lat}}$ was calculated by adding the plastic-wave transit time to this depth to $t_{\mathrm{ent}}$.

Stage~3 begins at $t_{\mathrm{lat}}$, when the modeled central peak-compression region closes. The elastic precursor subsequently reaches the rear free surface at $t_{\mathrm{rear,e}}$, calculated from the diamond thickness and elastic wave speed. This arrival launches a release wave that propagates toward the driven surface. The returning release front was assigned speed $c_e$, while the plastic front continued to advance at $D_p$. Their intersection defines $t_{\mathrm{int}}$ and the beginning of Stage~4.

\paragraph*{$[100]$ single crystal.}

The elastic and plastic wave speeds measured from the present radiography dataset were $c_e=17.32~\mu\mathrm{m/ns}$ and $D_p=14.01~\mu\mathrm{m/ns}$. These values give a diamond erosion angle of $\gamma_2^{[100]}\approx36.0^\circ$ and a centerline intersection depth of $d_1^{[100]}\approx110.1~\mu\mathrm{m}$. The corresponding lateral-release time is $t_{\mathrm{lat}}^{[100]}\approx11.9~\mathrm{ns}$. For the adopted diamond thickness, the elastic precursor reaches the rear free surface at $t_{\mathrm{rear,e}}^{[100]}\approx18.4~\mathrm{ns}$, and the returning release front intersects the plastic wavefront at $t_{\mathrm{int}}^{[100]}\approx20.0~\mathrm{ns}$. Stage~3 therefore extends from approximately 11.9 to 20.0~ns, with rear-surface release launched at 18.4~ns.

\paragraph*{$[110]$ single crystal.}

The $[110]$ estimates used $c_e=20.4~\mu\mathrm{m/ns}$ and $D_p=11.9~\mu\mathrm{m/ns}$ from the radiography analysis reported by Katagiri \emph{et al.}~\cite{SM-Katagiri2023Transonic}. These values give $\gamma_2^{[110]}\approx54.3^\circ$ and $d_1^{[110]}\approx57.5~\mu\mathrm{m}$, corresponding to $t_{\mathrm{lat}}^{[110]}\approx8.8~\mathrm{ns}$. Using the same diamond thickness, the elastic precursor reaches the rear free surface at $t_{\mathrm{rear,e}}^{[110]}\approx16.3~\mathrm{ns}$, and the returning release front intersects the plastic wavefront at $t_{\mathrm{int}}^{[110]}\approx19.5~\mathrm{ns}$. Stage~3 therefore extends from approximately 8.8 to 19.5~ns, with rear-surface release launched at 16.3~ns.

The resulting stage intervals are summarized in Table~\ref{tab:timing-stages}. The first clearly resolved linear features in the $[100]$ delay sequence at 8~ns fall within Stage~2. At 12~ns, the $[110]$ rotation series samples Stage~3, while the $[100]$ series is near the Stage~2--3 transition.

\begin{table}[t]
    \centering
    \caption{
    Approximate time intervals for the four loading stages.
    All delays are referenced to the arrival of the drive laser
    at the front surface of the PP ablator.
    }
    \label{tab:timing-stages}
    \begin{tabular}{lcc}
        \hline
        Stage & $[100]$ interval (ns) & $[110]$ interval (ns) \\
        \hline
        Stage~1 & $0 \leq t < 4.0$     & $0 \leq t < 4.0$    \\
        Stage~2 & $4.0 \leq t < 11.9$  & $4.0 \leq t < 8.8$  \\
        Stage~3 & $11.9 \leq t < 20.0$ & $8.8 \leq t < 19.5$ \\
        Stage~4 & $t \geq 20.0$        & $t \geq 19.5$       \\
        \hline
    \end{tabular}
\end{table}

\section{Determination of elastic and plastic wavefront positions in MD}
\label{sec:md-wavefront-positions}

Elastic and plastic wavefront positions were determined from spatial stress profiles along the shock-propagation direction $z$. Compressive stresses were taken as positive. The longitudinal stress $P_{zz}$ and mean transverse stress $P_T$, defined as the mean of $P_{xx}$ and $P_{yy}$, are shown for the 6~ps configuration in Fig.~3(b). The maximum shear stress $\tau_{\max}$ was calculated in each spatial bin as half the difference between the largest and smallest eigenvalues of the local compressive stress tensor $\mathbf{P}$.

Wavefront positions were identified by spatial-gradient analysis with Savitzky--Golay smoothing. With $z$ increasing in the shock-propagation direction, the elastic front was located at the steepest decrease in $P_{zz}$ at the leading boundary of the shocked material. The plastic front was located at the steepest increase in $\tau_{\max}$ between the plastically compressed region and the region compressed by the elastic precursor. For the 6~ps configuration, the plastic-front position also coincided with the steepest decrease in $P_T$ shown in Fig.~3(b).

The elastic and plastic wave velocities were obtained from separate linear fits of the corresponding front positions as functions of time. The elastic-front positions were determined from $P_{zz}$, and the plastic-front positions from $\tau_{\max}$. The fitted slopes gave the respective wave velocities.

\section{Deformation-corrected crystallographic projection references}
\label{sec:deformation-corrected-projection-references}

Crystallographic projection references for the $[110]$ rotation series were calculated using local deformation gradients from the plastically compressed region of the 6~ps MD configuration (step~12000). Atoms were matched to the undeformed step-0 reference by Particle Identifier. Local deformation gradients were calculated using OVITO's Atomic Strain modifier with a 3.0~\AA{} neighbor cutoff. Reference crystallographic directions were transformed by the local deformation gradient, and plane normals by its inverse transpose.

At $\phi=0^\circ$, the beam direction, detector horizontal, and detector vertical were parallel to $[\bar{1}10]$, $[00\bar{1}]$, and $[110]$, respectively. The deformed directions and plane normals were rotated by $+\phi$ about $[110]$ using the right-hand convention, with the beam and detector fixed. Material-line angles were calculated by projecting the transformed directions onto the detector plane. Plane-trace angles were calculated from the intersections of the transformed planes with the detector plane.

Plane-trace references were calculated for $(11\bar{1})$ and $(111)$.
In-plane line references were calculated for $[112]$ and $[101]$ within
$(11\bar{1})$, and for $[11\bar{2}]$ and $[01\bar{1}]$ within $(111)$.
At each rotation angle, each reference curve reports the median of the
projected angles calculated individually for the retained atoms.

\section{Geometric projection of finite sheet models}
\label{sec:sheet-projection-method}

The models in Fig.~8(b) are unit-density ellipsoids
whose shortest axis is normal to the $(111)$ plane. At the reference
orientation, $\phi=0^\circ$, the detector axes are
$x\parallel[00\bar{1}]$ and $z\parallel[110]$, and the line of sight
is $y\parallel[\bar{1}10]$, so the sheets are viewed edge-on.
The models are rotated about the $z$ axis by $\mathbf{R}_z(\phi)$,
with positive $\phi$ rotating $+x$ toward $+y$.

Each ellipsoid is centered at the origin, with semiaxes $L$, $W$,
and $T$ along the orthonormal directions
$\hat{\boldsymbol{\ell}}$ (in-plane long axis),
$\hat{\mathbf{w}}$ (in-plane transverse axis), and
$\hat{\mathbf{n}}$ (plane normal), all expressed in detector
coordinates:
\begin{equation}
    q_0(\mathbf{r}) =
    \begin{cases}
        1, &
        \displaystyle
        \frac{(\mathbf{r}\cdot\hat{\boldsymbol{\ell}})^2}{L^2}
        + \frac{(\mathbf{r}\cdot\hat{\mathbf{w}})^2}{W^2}
        + \frac{(\mathbf{r}\cdot\hat{\mathbf{n}})^2}{T^2}
        \leq 1, \\[6pt]
        0, & \text{otherwise}.
    \end{cases}
    \label{eq:sheet-model}
\end{equation}
Here, $\mathbf{r}=(x,y,z)$. All models use $L=1$ and $T=0.04$
in arbitrary units. The in-plane isotropic model has $W=1$;
the elongated models have $W=1/6$, with
$\hat{\boldsymbol{\ell}}\parallel[11\bar{2}]$ or $[01\bar{1}]$
and $\hat{\mathbf{w}}=\hat{\mathbf{n}}\times\hat{\boldsymbol{\ell}}$.

The projected image is the path length through the rotated
ellipsoid along the line of sight:
\begin{equation}
    G_\phi(x,z)
    = \int_{-\infty}^{\infty}
    q_0\!\left(\mathbf{R}_z(-\phi)\mathbf{r}\right)\,\mathrm{d}y.
    \label{eq:sheet-projection}
\end{equation}
Rotation preserves volume, so
$\int\!\!\int G_\phi\,\mathrm{d}x\,\mathrm{d}z=4\pi LWT/3$
at every $\phi$, and any change in peak height reflects a
redistribution of this fixed total. To preserve changes in
peak height, the projections of each morphology at
$\phi=0^\circ$, $20^\circ$, and $40^\circ$ are normalized
to the peak of its edge-on projection:
\begin{equation}
    I_{\mathrm{rel}}(x,z;\phi)
    = \frac{G_\phi(x,z)}{\max_{x,z}G_0(x,z)}.
    \label{eq:sheet-projection-normalization}
\end{equation}

The overlaid reference lines mark the plane trace, parallel to
$\hat{\mathbf{n}}_\phi\times\hat{\mathbf{y}}$, and the projected
long axis,
$\hat{\boldsymbol{\ell}}_\phi
-(\hat{\boldsymbol{\ell}}_\phi\cdot\hat{\mathbf{y}})
\hat{\mathbf{y}}$.
Here, $\hat{\mathbf{y}}$ is the unit vector along the line of sight,
$\hat{\mathbf{n}}_\phi=\mathbf{R}_z(\phi)\hat{\mathbf{n}}$, and
$\hat{\boldsymbol{\ell}}_\phi
=\mathbf{R}_z(\phi)\hat{\boldsymbol{\ell}}$.

\section{Fourier visibility}
\label{sec:fourier_visibility}

Fourier visibility was evaluated from fixed regions in the experimental radiographs and MD defect-fraction projections. After background subtraction and windowing, the Fourier power spectra were radially whitened.

For both datasets, signal windows were centered on the Fourier-normal directions predicted by the ideal-crystal $[112]$ and $[11\bar{2}]$ projections during rotation about $[110]$:
\[
\beta_{\pm}(\phi)
=\pm\arctan\!\left(\frac{1}{\sqrt{2}\cos\phi}\right),
\qquad
\theta_{\pm}(\phi)
=[\beta_{\pm}(\phi)+90^\circ]\bmod180^\circ.
\]
Here, $\beta_{\pm}$ are the projected feature angles relative to the detector horizontal.

For each family $j$, the mean whitened power in the signal window, $S_j$, was compared with the average power in two flanking background windows, $B_j$, over the same radial band:
\[
V_j=\frac{S_j-B_j}{S_j+B_j}.
\]
The two family visibilities were averaged at each rotation angle. Each dataset's mean visibility was then normalized to its value at $\phi=0^\circ$ to obtain $V_{\mathrm{rel}}(\phi)$.

\clearpage
\bibliography{references_supplement}

\clearpage
\begin{figure}[!ht]
    \centering
    \includegraphics[width=1.0\textwidth]{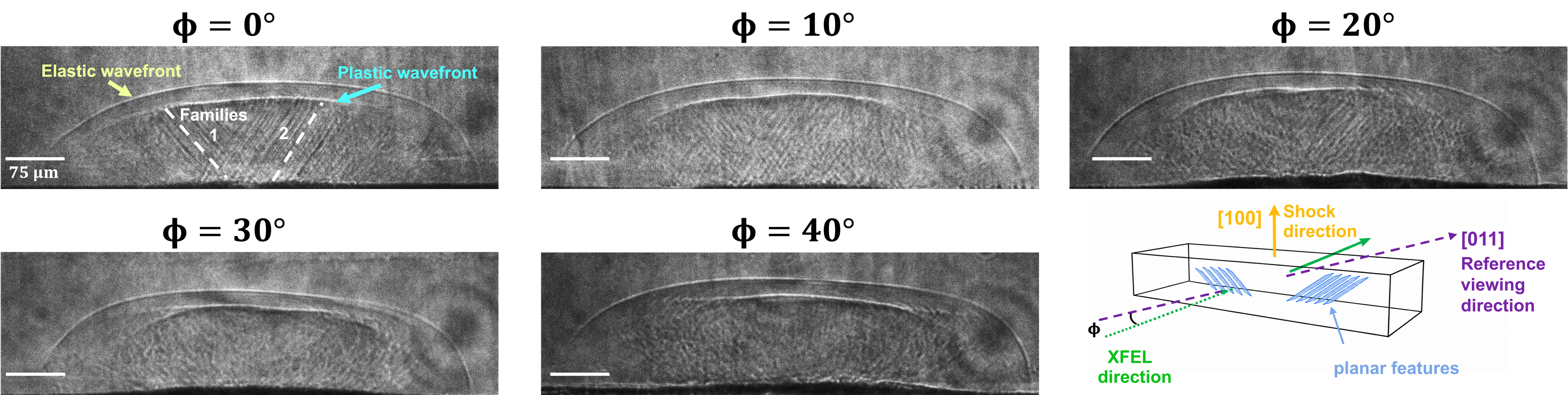}
    \caption{\textbf{Viewing-direction dependence of linear contrast in single-crystal diamond shocked along $[100]$.} Selected XFEL radiographs acquired at a pump-probe delay of 12~ns are shown for crystal rotations about the fixed $[100]$ shock axis. 
    In the $\phi=0^\circ$ radiograph, light-yellow and cyan arrows mark the elastic and plastic wavefronts, respectively; white dashed guides mark Families~1 and~2. 
    The schematic shows the viewing geometry, with internal features represented as planes. The rotation angle $\phi$ is measured from the reference XFEL viewing direction $[011]$. 
    Scale bars, 75~$\mu$m.}
    \label{fig:figS1}
\end{figure}

\clearpage

\begin{figure}[!ht]
    \centering
    \includegraphics[width=1.0\textwidth]{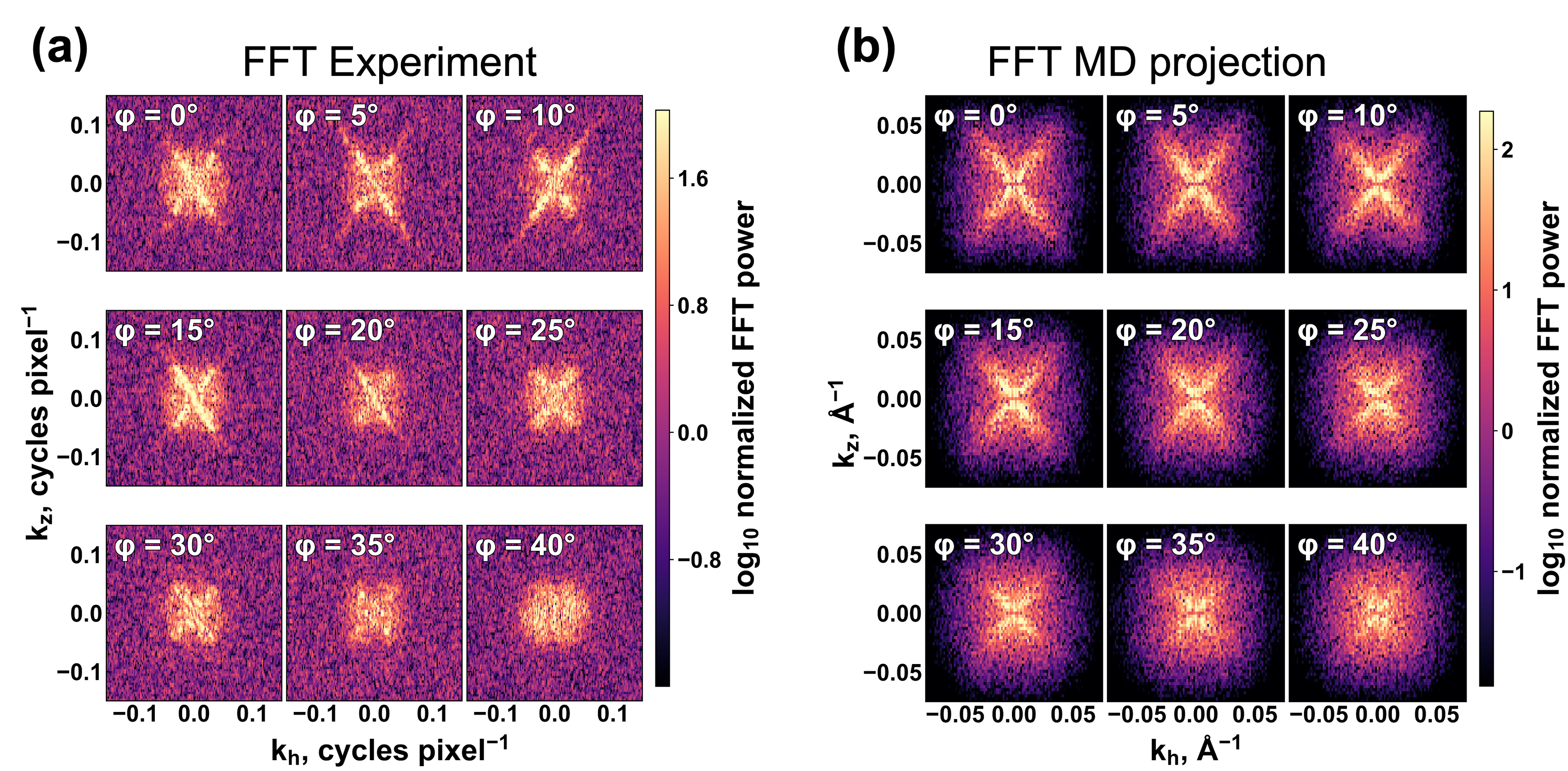}
    \caption{\textbf{Full rotation series of Fourier power spectra for experimental radiographs and MD defect projections.}
    (\textbf{a}) Two-dimensional Fourier power spectra of the experimental radiographs acquired at a pump--probe delay of 12~ns.
    (\textbf{b}) Corresponding Fourier power spectra of defect projections calculated from the 6~ps MD configuration. 
	Spectra are shown for $\phi=0^\circ$--$40^\circ$ in $5^\circ$ increments, where $\phi$ denotes rotation about the $[110]$ shock axis and $\phi=0^\circ$ corresponds to viewing along $[\bar{1}10]$. The horizontal and vertical spatial-frequency components, $k_{\mathrm{h}}$ and $k_z$, are expressed in $\mathrm{cycles\,pixel^{-1}}$ in (\textbf{a}) and $\text{\AA}^{-1}$ in (\textbf{b}). Colors indicate the logarithm of the normalized Fourier power. The spectra at $\phi=0^\circ$, $20^\circ$, and $40^\circ$ are also shown in Fig.~4(b,d).}
    \label{fig:Sup_ExpMDFFT}
\end{figure}

\clearpage
\begin{figure}[!ht]
    \centering
    \includegraphics[width=1.0\textwidth]{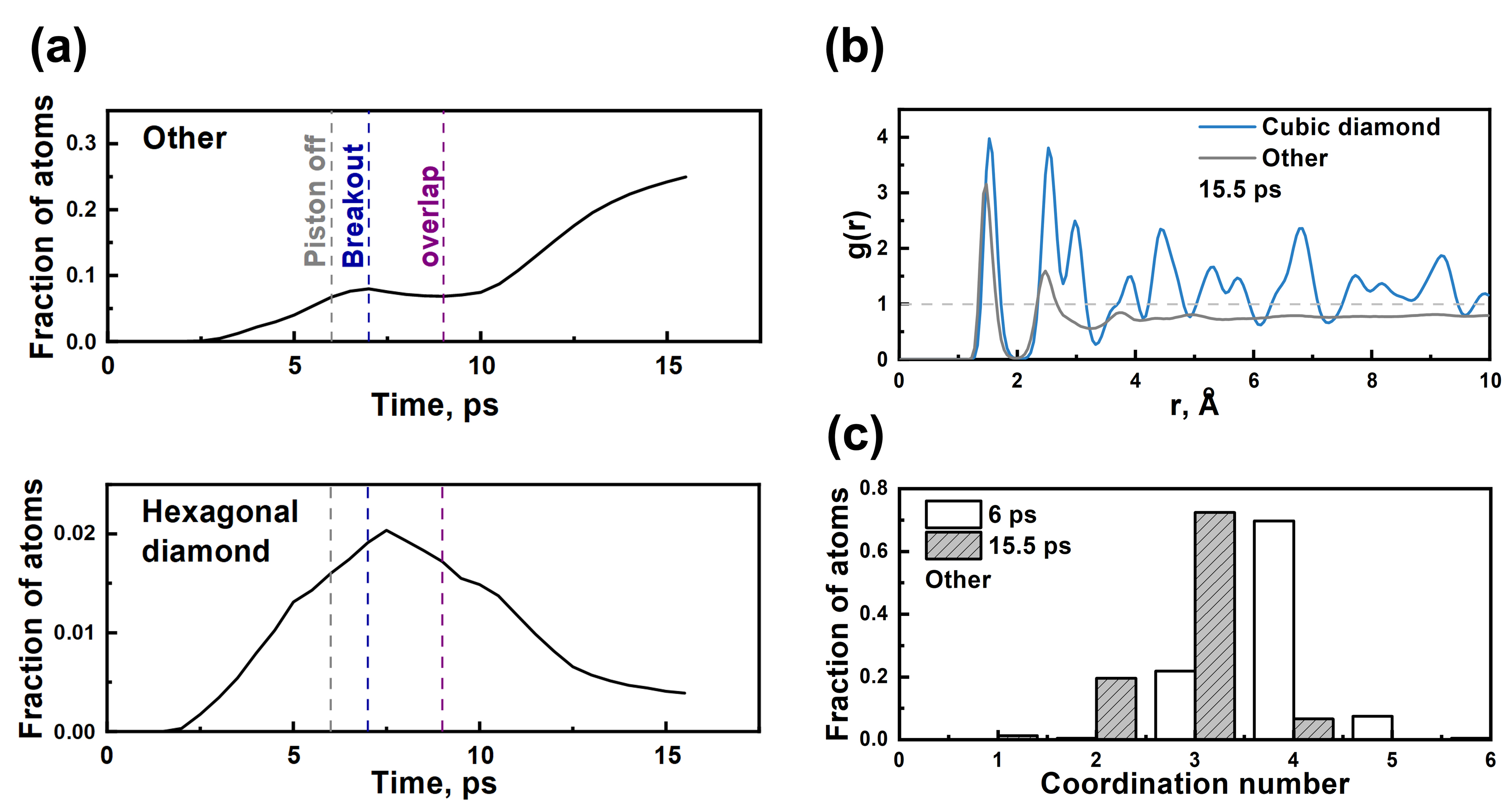}
	\caption{\textbf{Evolution of local atomic environments within
the simulated defect network during compression and release.}
	(\textbf{a}) Fractions of atoms classified as \textit{Other}
(top) and Hexagonal diamond (bottom) within the ROI shown in Fig.~6(a), as functions of simulation time. Vertical dashed lines mark piston shutoff (gray), elastic-wave breakout at the upper free surface (blue), and the encounter of the returning release front with the plastic
front (purple).
	(\textbf{b}) Radial distribution functions of the cubic-diamond
and \textit{Other} populations within the same ROI at 15.5~ps. The horizontal dashed line marks $g(r)=1$.
    (\textbf{c}) Coordination-number distributions of atoms classified as \textit{Other} at 6~ps (unfilled bars) and 15.5~ps (hatched bars), normalized by the number of \textit{Other} atoms at each time.}
    \label{fig:Sup_ROI}
\end{figure}

\clearpage